\documentstyle[12pt,aasms4]{article}
\def\iras{{\sl IRAS\/}}

\def\msol{M$_\odot$}
\def\mloss{M$_\odot$ yr$^{-1}$}

\begin{document}
 
\title{
A Mid-Infrared Imaging Survey of Proto-Planetary Nebula Candidates
}

\author{Margaret Meixner and Toshiya Ueta}
\affil{Dept. of Astronomy, MC-221, University of Illinois, Urbana, IL  61801,
meixner@astro.uiuc.edu, ueta@astro.uiuc.edu}

\author{Aditya Dayal}
\affil{IPAC/JPL, Caltech, MS 100-22,
770 South Wilson Ave. Pasadena, CA
91125, adayal@ipac.caltech.edu}

\author{Joseph L. Hora and Giovanni Fazio}
\affil{Harvard-Smithsonian Center for Astrophysics, MS 65,
60 Garden St., Cambridge, MA  02138,
jhora@cfa.harvard.edu, gfazio@cfa.harvard.edu}

\author{Bruce J. Hrivnak}
\affil{Dept. of Physics \& Astronomy, Valparaiso University,
Valparaiso, IN  46383 bhrivnak@exodus.valpo.edu}

\author{Christopher J. Skinner}
\affil{deceased}

\author{William F. Hoffmann}
\affil{Univ. of Arizona, Steward Obs., Tucson, Az  85721, whoffmann@as.arizona.edu}

\author{Lynne K. Deutsch}
\affil{Boston Univ., Dept. of Astronomy/CAS 519, 725 Commonwealth Avenue, Boston, MA,  02215,
deutschl@bu.edu}

\accepted{to \apjs ~May 1999 issue}

\begin{abstract}

We present the data from a mid-infrared imaging survey of 66 proto-planetary
nebula candidates using two mid-IR cameras (MIRAC2 and Berkcam)
at the NASA Infrared Telescope Facility and the United Kingdom Infrared
Telescope.  The goal of this survey is to determine the size, flux,
and morphology of the mid-IR emission regions, which sample the inner
regions of the circumstellar dust shells of proto-planetary nebulae.  
We imaged these proto-planetary nebulae with narrow-band filters
($\Delta\lambda / \lambda \sim 10\%$) at wavelengths of notable
dust features. 
With our typical angular resolution of 1\arcsec, we resolve 17 sources,
find 48 objects unresolved, and do not detect 1 source.
For several sources, we checked optical and infrared associations and
positions of the sources. In table format, we list the size and flux
measurements for all the detected objects and show figures of all the
resolved sources.   Images for all the detected objects are available on line
in FITS format from the Astronomy
Digital Image Library at the National Center for Supercomputing
Application.   The proto-planetary nebula candidate
sample  includes, in addition to the predominant proto-planetary nebulae,
extreme asymptotic giant branch stars, young  planetary
nebulae, a supergiant, and a luminous blue variable.
We find that dust shells which are cooler ($\rm T \sim 150$ K) and 
brighter in the infrared are more easily resolved.
  Eleven of the seventeen resolved sources are extended
and fall into one of two types of mid-IR morphological classes: core/elliptical
or toroidal.  Core/elliptical structures show unresolved cores with lower
surface brightness elliptical nebulae.  Toroidal structures show 
limb-brightened peaks suggesting equatorial density enhancements.  We argue
that core/ellipticals have denser dust shells than toroidals.

\end{abstract}


\section{Introduction}

Proto-Planetary Nebulae (PPNe) are objects in transition between the
asymptotic giant branch (AGB) and planetary nebula (PN) phases of
stellar evolution for intermediate mass stars  ($\sim$ 0.8$-$8.0 \msol).
Like PNe, PPNe are composed of central stars and detached
circumstellar envelopes of gas and
dust; however, unlike PNe,
their central stars are too cool ($\rm T_{eff} < 10^4$ K) 
to photoionize the envelopes (\cite{kwok93}).  
Hence, we cannot
observe the structure of PPNe circumstellar envelopes using 
high spatial resolution imaging of ionized gas lines or free-free radio
continuum that are successfully used for PNe.  
Instead, techniques which are sensitive to dust 
(e.g. thermal emission at infrared wavelengths 
or scattering at visible wavelengths) or 
neutral gas (e.g. CO or OH mapping in radio) must be used.  

In this paper, we present results of a mid-infrared  
(mid-IR; 8$-$25 \micron)
imaging survey of 66 PPN candidates.  These images are sensitive
to the thermal dust emission arising from the inner-most regions
of the PPNe circumstellar dust shells.
Figure \ref{cartoon} shows our working model for
the circumstellar dust shell
created by the AGB and superwind mass loss  episodes (cf. \cite{meixner97}).   
If the mass loss velocity is relatively constant
over time, then an increasing radial distance in the dust shell corresponds to
a time further in the past.  
The outer regions of the dust shell consist of material
that was ejected further in the past than material in the inner regions.
The inner radius of the dust shell
marks the point in  time when the mass loss ended. 
The ``superwind'' region marks the location
of material ejected during an intense and perhaps axially symmetric
mass loss phase that ends a star's life on the AGB. The central
star heats the dust shell directly and, hence, the
temperature of the dust decreases with increasing radial distance from the 
central star. 
Judging from the spectral energy distributions of many PPNe, the temperature
of the dust at the inner radius is about a few 100 K  (\cite{volk89b}).
Warm dust  (i.e. at a few 100 K) emits  primarily  in the mid-IR. 
Hence, the mid-IR emission arises from
the inner radius out into the superwind region to a radial distance 
where the dust temperature drops below a 100 K  
(shaded region of Fig. \ref{cartoon}; \cite{meixner97}). 
The purpose of this survey is to determine the size
and morphology of the mid-IR emitting regions.  The size  of this region
gives us direct information on the inner radius of the dust shell 
and the extent
of the warm  dust.  The morphology of this region shows the geometry
of the superwind phase.  

We selected our sample of 66 PPN candidates from  many articles in the 
literature and our personal knowledge of the sources.  
To provide a context for our discussions below,  we have created Table
1, which lists some pertinent characteristics  of each  these PPN 
candidates from the literature.  Included in Table 1 is the
source's  \iras~ID, its chemistry (C = carbon-rich/C-rich, 
O = oxygen-rich/O-rich, C/O = both C-rich and O-rich), 
the spectral type of the central star, the visual
magnitude of this central star, the \iras~fluxes in Jy, the reference
for a mid-IR spectrum, a distance estimate, and references
for the listed information.   When information is not listed, it means
that it could not be  found.  A brief look at Table 1 shows the breadth
of the sample of PPNe that we imaged in this survey: an almost even
mix of C- and O-rich sources, a spread of spectral types from M to O,
and  a range of infrared brightnesses ($\rm F_{25} = 2.4 - 2314$ Jy) 
and visual magnitudes (24$-$5.5). In Roman numeral footnotes,
we note  14 of the sources which
are not strictly PPNe, but very young planetary nebulae (Young PN), 
extreme AGB stars\footnote{An extreme AGB star is a heavily dust 
enshrouded star
in the process of its final mass loss episode. Extreme AGB stars are
at the so-called tip of the AGB.}, or other.
For brevity, we refer to this entire sample as PPNe or PPN candidates
hereafter.

The remainder of this paper covers  the observations (section 2),
the results (section 3), a brief discussion of the results (section 4),
and the conclusions (section 5).  However, the bulk of the information
can be found in Tables 1, 2, \& 3 and the figures.

\section{Observations and  Data  Reduction}

Observations of the PPN candidates were obtained using two different
mid-IR cameras, MIRAC2 and Berkcam, between 1991 and 1995.
The specific observing parameters  for each  PPNe  are listed  in Table 2.
In particular,  the positions  recorded in this table were the positions
at which we found the sources and are accurate within a few arcseconds.
The angular resolution, which can be assessed for individual images
by looking at the point spread function (PSF) size in Table 3, is 
typically about 1\arcsec.
Diffraction effects dominate the angular resolution for many of the images.
For a 3 m telescope, the diffraction-limited angular resolution is
$\sim 0$.\arcsec 8 at 10 \micron.  
Below we describe the general observation strategy used with  the MIRAC2
and the Berkcam  mid-IR cameras.

\subsection{MIRAC2}

We obtained images  of 52 PPN candidates
using MIRAC2, the University of Arizona/SAO
mid-IR  camera  (\cite{hoffmann98}),  at the 3.0 m NASA Infrared 
Telescope Facility
(IRTF) and at the 3.8 m United Kingdom  Infrared  Telescope  (UKIRT).  The
dates of the observations  span three observing runs: 
(1)  1995 June 3$-$7 at IRTF,  
(2)  1995 November 6, 7, 10 at IRTF, and 
(3)  1995 November 22$-$24, 28 at UKIRT.  The weather was
clear and calm for the entire 1995 June run, was mostly clear for
the 1995 Nov IRTF run and was mostly cloudy for the 1995 Nov 
UKIRT run.  The array is a Boeing HF-16 arsenic-doped silicon 
blocked-impurity-band
hybrid array and has a $128 \times 128$ pixel format.  
The pixel scale,  which can be changed with a zoom
magnification,  was set to 0.\arcsec 39  for the 1995 June run,
to 0.\arcsec 34 for the 1995 Nov IRTF run, and 0.\arcsec 28 for the
1995 Nov UKIRT run.  These pixel scales ensure  a Nyquist sampling
of the diffraction-limited  PSF at both telescopes. 
The fields of view were 
50\arcsec$\times$50\arcsec, 
44\arcsec$\times$44\arcsec, and  
36\arcsec$\times$36\arcsec, respectively.   
Wavelengths of observations are selected  using  
interference filters (bandwidths listed in Table 2) or 
circular variable filters (labelled as CVF with corresponding
bandwidth in Table 2).  The strategy
was to image these PPNe at wavelengths of known dust features in the mid-IR 
and  at wavelengths of the dust continuum.  Mid-IR spectra, when 
available,  were used for guidance in selecting wavelengths.  We
used the standard method of chopping and  nodding to sample the sky 
background emission in the mid-IR. 
The secondary was chopped by 20\arcsec~south at a $\rm \sim 3$ Hz rate 
and the telescope was nodded by 20\arcsec~west at a $\rm \sim 0.07$ Hz rate.
For \iras~18184-1623, which is more extended, the chop and nod throws
were increased to  40\arcsec.
The CGS3 standard stars (cf. \cite{cohen95}) were used 
for flux and PSF calibration. 
We observed standard stars before and
after the sources  to check for variations in the PSF. 
For high declination
sources at the IRTF, we used 
K giants  near the source (within a few degrees) as PSF standards  
because there are no flux standards at these high declinations.  
This additional PSF check 
is particularly necessary because the IRTF has an astigmatism 
which affects the PSF, especially at high declinations.

For each observed wavelength of a source, 
we obtained $\sim$10 sets of nod-chop cycles where
each nod-chop cycle consists of 2 nod sets  (1 on; 1 off) and each
nod set contains 2 chopped frames  (1 on; 1 off) for a total of 4
frames for a nod-chop cycle.  Using our own software, we first
subtracted the chop-off from the chop-on frames   
and then subtracted the nod-off  from the nod-on frames;
and thereby, obtained sky-subtracted, co-added images
in which the source appears in all four beams (two positive and two negative). 
A  gain matrix to flat field the images was derived  from images of 
the dome  (high intensity, uniform background)  and the sky  (low intensity,
uniform background) and applied  to the sky-subtracted, co-added images. 
A mask file to block out the effects of bad pixels and field
vignetting was created and applied to these flat-fielded images.
The images were further processed using IRAF\footnote{Image Reduction 
and Analysis Facility is distributed by the National
Optical Astronomy Observatories, which is operated by the Association of
Universities for Research in Astronomy, Inc., under cooperative agreement with
the National Science Foundation.} routines.
In order to cross correlate the multiple sky-subtracted, co-added images on 
a sub pixel scale,
we first  magnified the images by a factor of four, using a linear
interpolation and conserving the flux.  The resulting images had  pixel
sizes 1/4 of their original size:
0.\arcsec098/pixel for the 1995 June run, 
0.\arcsec085/pixel for  the 1995 Nov IRTF run, and 
0.\arcsec070/pixel for the 1995 Nov UKIRT run.
The multiple sky-subtracted, co-added images of each source 
were cross-correlated, realigned, 
summed, and dissected into  four individual images of the source.
These four images were cross-correlated, realigned and 
summed to produce the final image.  
The same observing and reduction procedures
were used for the standard  stars and the sources;  thus any
effect of these procedures on the resultant images of the sources and 
standards should be the same.
Typical integration times  on source are 70 seconds,
resulting in typical one sigma, rms noise  of  1.1 mJy arcsecond$^{-2}$.

\subsection{Berkcam}
 
Observations and data reduction 
with Berkcam followed a similar strategy as observations
with MIRAC2 and we describe the  main differences here.
We observed  14 PPN  candidates using Berkcam, which had a 
10$\times$64 Hughes  Aircraft, Si:As, RIBIT (Reversed Illuminated
Blocked Impurity Transmission) array (\cite{arens87}; \cite{keto92}).
The observations  were made
in 1992 June and 1992 December at UKIRT  and  in 1991 May at IRTF.
The weather on all nights of observations was clear.
The pixel scale was 0.\arcsec 39.  The resulting small field
of view (4\arcsec$\times$25\arcsec) of the array
required us to  mosaic images of all sources and standard stars
with 3$-$5 array panels in order to cover the entire image of the object.  
Chopping the secondary with a 10\arcsec$-$40\arcsec~throw at a rate of
 6.5 Hz and nodding the telescope with a 20\arcsec$-$40\arcsec~throw
every 5$-$15 seconds was performed to measure the sky background
(cf. \cite{ball92}).
 The standard stars  were
observed  before and after each of the sources.  The differences  in airmasses
between the standard stars and sources  were small ($< 0.2$). 

We used IRAF  for the initial stages of data reduction.  
We first obtained  sky-subtracted, co-added images, similar to those we 
obtained for the MIRAC data,  except that a partial 
source appears once, sometimes
twice on the final image depending on the size of the chop and nod throws.
We derived flat fields directly from sky frames from which we subtract
dark frames and applied them to the sky-subtracted, co-added images.  
In order to stitch together the 3-5 panel mosaics of
images on a subpixel scale, 
we first  magnified by a factor of four, resulting in a pixel
size of 0.\arcsec098/pixel. 
The sky-subtracted, flat-fielded, magnified
mosaic panels  were then stitched together   
using an automatic mosaicing program, which we developed and that
determines the correct relative offsets for the panels  using a minimum
least-squares error in the flux differences  as a criterion.  
 Typical one sigma
rms noise of the images is 46 mJy arcsec$^{-2}$.

\subsection{ Flux calibration}

We flux-calibrated our source by comparing the source counts (analog to
digital units; ADUs)  with the standard star counts. 
Extinction corrections were negligible and not applied because the 
airmass differences 
between the standard star and source were kept to less than 0.2.
The conversion from ADU to Janskys is done by the following equation;
\begin{equation}
\rm
        F_{\nu}^{src}(\mbox{Jy}) 
        = \frac{F_{\nu}^{src}(\mbox{ADU})}{F_{\nu}^{std}(\mbox{ADU})} 
        \times F_{\nu}^{std}(\mbox{Jy}),
        \label{fluxconv}
\end{equation}
where $F_{\nu}^{src}$ ($F_{\nu}^{std}$) is the source (standard)
flux density at the observed wavelength 
($\rm \lambda = {\nu \over c}$).  
The flux densities for  the standards 
are from the CGS3 Standard Table and are specified
for wavelengths of 8.65, 10, 11.5 and 19.3 \micron~(cf. \cite{cohen95}).
For many of the wavelengths,  we needed to interpolate values from the
table and we assumed a power-law relation between flux and wavelength:
\begin{equation}
\rm
        F_{\nu}^{std}(\mbox{Jy}) 
            = F_{\nu_{0}}^{std}(\mbox{Jy}) \left( \frac{\lambda}{\lambda_{0}} 
            \right)^{\beta}, 
        \label{p-law}
\end{equation}
where $\lambda$ ($\lambda_{0}$) is observed (tabulated) wavelength 
and $\beta$ is the power-law coefficient derived from our interpolation
between two values of the CGS3 Standard 
Table.   The values for $\beta$  average -1.9$\pm$0.1  for all the stars, 
in the wavelength ranges 10-11.5 and 11.5-19.3 \micron, which is close to 
-2, the value expected for the Raleigh-Jeans approximation in this wavelength 
range. The value for $\beta$ in the wavelength range, 8.65-10 \micron~for those
stars with an SiO absorption band is more shallow, an average of -1.6$\pm$0.2, 
because the SiO absorption band at 8 \micron~causes an apparent rise in
flux between 8 and 10 \micron~relative to the expected -2 fall off 
(\cite{cohen95}).
Absolute flux calibration errors are at least 10\%.
Photometry of the 1995 Nov UKIRT data was not reliable due to poor
weather conditions, and is not
reported  in Table 3.

\section{Results}

Of the 66 PPN candidates, 17 (26\%) are resolved, 48 are unresolved and 
1 is undetected.
In this section, we present  the sizes and fluxes  of
the mid-IR emission region of the  65 detected PPN candidates and discuss
the images of two unresolved sources and all seventeen resolved sources. 
In section 3.1, we describe our method for measuring the sizes and fluxes
listed in Table 3.  In section 3.2, we talk about optical associations
for the infrared components of some PPN candidates and the undetected
source. In section 3.3,
we describe how we determined whether or not our mid-IR images resolve
the sources.  The cross cuts of three resolved sources and their PSFs
(Figure \ref{figslice}) illustrate three categories:  marginally resolved,
core/elliptical and toroidal.  In sections 3.4 and 3.5, we discuss, 
respectively, the 
marginally resolved sources and extended sources, of which core/ellipticals
and toroidals are two distinct subcategories.

\subsection{Measurements \label{measure.}}

In Table 3, we list the measured properties of the sources for each 
observed wavelength: flux density, size and ellipticity of source, and, for
comparison, size and ellipticity of the PSF. Most objects have at least
two PSFs (before and after) for comparison. We obtain these measurements 
by fitting isophotal ellipses to  the infrared emission.  Total flux
density is summed within the $3\sigma$ isophote.
From the isophote 
with 50\% of the peak intensity, we measure the  size (Size), 
ellipticity (EL), and position angle (PA). 
Size is the major axis length in arcseconds.
EL is defined as $\mbox{EL} = 1 - b/a$, 
where $a$ ($b$) is the major (minor) axis length.
PA is measured in degrees 
from north toward east (counter-clockwise in the sky).

\subsection{Optical Associations and Non-detections}

A number of these PPNe have supposed optical associations established
by either cross correlating optical star catalogs and the \iras~catalog 
or by looking for optical counterparts on the all sky survey within the 
\iras~error ellipse.
During this mid-IR imaging survey, we found that some of these optical
associations are correct  and others  are not.   We determined the validity
of an association by finding the optical counterpart in the telescope
optical guide camera, and checking the position of the corresponding mid-IR
image  on the array.  The alignment
between the optical guide camera and the mid-IR camera was established
by imaging standard stars which are bright at optical and mid-IR 
wavelengths.  
We confirmed the optical associations for  \iras~00470+6429, 
\iras~04386+5722,
and \iras~19386+0155. \iras~22480+6002 had two possible optical
associations, both 
bright stars  separated by
10\arcsec~east-west, which fall
within the position error ellipse.  We found that \iras~22480+6002
is associated with  the brighter star that lies west in the pair.
We also found that HD 235718 (SAO 34043) is not associated with 
\iras~22036+5306, and that there is no infrared counterpart to HD 235718.
We do not detect mid-IR emission from  HD 20041 which \cite{oudmaijer92}
claim is associated with  \iras~03119+5657.

\subsection{Criteria for Resolving Sources} 

We determine if a source is extended, marginally resolved or
unresolved by systematically comparing the source
size with its PSF size  in two ways: a visual comparison of the
source and PSF images and cross cuts (e.g. Fig.\ref{figslice}),  and a 
quantitative comparison of the source and PSF sizes listed in Table 3.  
For the quantitative comparison,
we calculate  a source-to-PSF size ratio  for each wavelength, and then 
obtain an average ratio for all observed wavelengths.  
For reference, we also calculate a [larger PSF]-to-[smaller PSF]
ratio from the  before and after
source observations of PSFs and find an average ratio of 1.08$\pm$0.1 for 
all the PSFs  and an  average ratio of 1.06$\pm$0.07 if we exclude UKIRT data. 
We claim that a source is  extended 
if this averaged source-to-PSF size ratio is larger  than 1.3 and 
unresolved if the averaged source-to-PSF size ratio is
smaller than 1.1. Sources that have
averaged source-to-PSF size ratios  between 1.1 and 1.3  are considered
marginally resolved, depending  on the quality and quantity of the data.  
For example,  several sources which were 
observed during the 1995 Nov UKIRT
run fall into this 1.1 to 1.3  range; however, we do not think any of them
are resolved  because of the variable seeing during that run.  
The IRTF images were taken with 
much better seeing  but some of them are affected by  astigmatism (see below).
Sources with ratios between 1.1 and 1.3  at more than
one wavelength  (e.g. \iras~02229+6208) are 
considered marginally resolved, while others in this range,
for which we have data at only one wavelength 
(e.g. \iras~17534+2603), are unresolved.
\iras~06176$-$1036 (Red Rectangle)
is an exception to the above quantitative criteria.  Visual
comparison of \iras~06176$-$1036  and its PSF shows that it is most certainly
extended at the 20\% contour level,  but quantitative comparison suggests
that it is unresolved because it is unresolved   at the 50\% contour
level used in our quantitative criteria. 

Using the above criteria,
we determine that a total of 17 sources are resolved; 11 are clearly
extended and 6 are marginally resolved. We summarize  information about the 
resolved  sources in Table 4A:
\iras~ID, mid-IR size, mid-IR  morphological class, chemistry, type of
object, and optical morphology.  Looking at Table 1,
where the general observed properties of these PPNe candidates are listed,
we ask the question: 
Are there any physical characteristics which determine whether
or not a source is resolvable?  Distance of the source, the luminosity
of the central star and the inner
radius of the dust shell are three important factors because if a 
source is closer it is more easily resolved, if it is more luminous
the size of the warm dust emission region is larger, and if the inner
radius of the dust shell is larger it is more easily  resolved.   
In Table 1, we list
distance (D) estimates  for some sources,  but these estimates are 
reliable only to a factor of a few and, hence, the luminosities (L) for the
sources are also not well known. However, if we compare the Table 1  
\iras~fluxes   of the sources, which are proportional to LD$^{-2}$, we find
that  sources  which
are resolved  tend to have higher fluxes,  and hence are 
closer or more luminous or more luminous and closer  than the
unresolved sources.  Inner radii of the dust shells
are not known; in fact these mid-IR images are the best sources  for
this information.  However,  the temperature of  the dust, as determined
from the infrared spectral energy distribution (SED), 
can yield an estimate on the inner radius.  Cooler ($\rm \sim 150 K$) 
dust shells, where
the SED peaks  at  25\micron, have larger inner radii than  warmer
($\rm > 300 K$) dust shells, where the SED peaks at 12\micron~or shorter 
wavelengths, because  the dust is cooler when it is 
further from the  central star which
heats it (\cite{volk89b}).   
All of the sources that are resolved, except \iras~06176-1306, 
are significantly brighter  ($\times$2-10) at 25\micron~compared to 12\micron,
while many of
the  unresolved  sources  are brighter at 12\micron~compared to 25\micron.
Thus we conclude that sources  which have brighter  and cooler infrared
fluxes are more easily resolved.   We add one speculative footnote to
this discussion.  The three faintest resolved sources, \iras~20000+3239,
\iras~22223+4327 and \iras~23304+6147, all have SEDs which peak at 
25\micron~but are not particularly bright compared to other sources in
Table 1.   All three of these sources are C-rich and two have dust
features attributed to polycyclic aromatic hydrocarbons (PAH) in their
mid-IR spectra (the third has no published mid-IR spectrum; cf. Table 1).  
We suggest
that the PAH dust emission in these sources tends to be more extended than
one would expect from dust grains in thermal equilibrium; thereby making
it easier to resolve these sources.

\subsection{Unresolved and Marginally Resolved Sources}
 
The images of two unresolved sources and all of 
the seventeen resolved sources are shown in Figure \ref{figmidirimag} 
and comments about each source appear in appendix A.
The MIRAC2 images  of \iras~07399$-$1435 and all of the sources imaged
with Berkcam  have been presented in a preliminary
form by \cite{dayal97} and \cite{meixner93a}, respectively.

The two unresolved sources, \iras~16559$-$2957  and \iras~22480+6002, 
show two extremes in the quality of PSFs.
The image of \iras~16559$-$2957  is a very good example of an
unresolved source: the source looks exactly like the PSFs. 
The PSFs for \iras~16559$-$2957
are also nicely round.
The images of \iras~22480+6002 and its PSFs are presented to
show how IRTF astigmatism affects the shape of the images, particularly
at high declinations.  \iras~22480+6002 has a cross  appearance: 
a main NE-SW elongation with
some perpendicularly extended structure. 
The standard star $\beta$ Peg, which is at a
significantly lower declination,  has a very round appearance, and thus 
one might
think that \iras~22480+6002 is extended.  However, the image of 
a K star nearby 
\iras~22480+6002 ($<$ 1$^\circ$ separation) has an appearance similar
to the image of \iras~22480+6002.
Thus, this elongation  is an artifact, caused by a known
astigmatism problem  with the IRTF.  Hence, we consider
such elongations and cross structures, particularly for high declination 
sources, also to be artifacts. \iras~23304+6147, discussed below, is an
example of a marginally extended source affected by this astigmatism.

The six  marginally-resolved sources are:
\iras~02229+6208, \iras~17441$-$2411,
\iras~18184$-$1302, \iras~18276$-$1431, \iras~20000+3239, and 
\iras~23304+6147
(Figure \ref{figmidirimag}).   The cross-cut of
\iras~02229+6208 displays a comparison of a marginally resolved source
with its PSF (Fig. \ref{figslice}).    
Three of these sources, \iras~02229+6208, \iras~20000+3239, and 
\iras~23304+6147,  
are cool ($\rm T\sim$150 K), C-rich PPNe with  features at 21 \micron.
Two  sources, \iras~17441$-$2411 and \iras~18276$-$1431, are  cool, O-rich
PPNe.   \iras~18276$-$1431 is marginally extended
at 9.8 \micron~but unresolved at 12.5 \micron.  The remaining source,
\iras~18184$-$1302, is probably a pre-main sequence star which was
included as a PPN candidate  because of its cool \iras~colors.

\subsection{Extended Sources:  Core/Elliptical  vs.  Toroidal}

Eleven sources are extended 
(source-to-PSF ratio $>$ 1.3),
ten of which appear to have an axial symmetry.
The axial symmetry
observed in the extended images seems to fall into two 
distinct morphological classes:
(1) core/elliptical and 
(2) toroidal (Table  4A).
The core/ellipticals  have  unresolved, very bright cores that are surrounded
by a lower surface brightness elliptical nebula. 
The toroidals have some evidence for two peaks  that are 
interpreted as limb-brightened peaks of an equatorial density enhancement.
Cross cuts through the images of the core/ellipticals and toroidals help
to illustrate these differences (Fig. \ref{figslice}).  
Core/ellipticals are strongly peaked with a broad
pedestal,  while toroidals have a plateau-like 
appearance when the two peaks are not
quite resolved.

\subsubsection{Core/ellipticals}

Eight of the sources  are core/ellipticals:  \iras~06176$-$1036 (Red Rectangle), 
\iras~07399$-$1435 (OH231.8+4.2),
\iras~16342$-$3814, \iras~17028$-$1004 (M2$-$9), 
\iras~17150$-$3224, \iras~17347$-$3139, \iras~19244$+$1115
(IRC+10420) and \iras~19374$+$2359 
(Figure \ref{figmidirimag}). 
All but one, \iras~06176$-$1036, are O-rich.
\iras~16342$-$3814  is a particularly striking core/elliptical source.
The  elliptical nebulae in \iras~16342$-$3814 extends NE-SW with
the same PA  for all wavelengths.  The cross-cut through the
major axis of this nebula reveals the contrast between the
unresolved core and the lower surface brightness nebula
(Fig. \ref{figslice}).   The apparent increase in this
elliptical nebula's size from the 12.5 \micron~image to the 20.6 \micron~image
is due largely to the telescope's diffraction effects which limit the
angular resolution of the images.  For diffraction limited images of
point sources, we expect the image size to scale proportionally with
the wavelength ($\lambda$); specifically, for images at 12.5 \micron~and
20.6 \micron~we expect an increase of 20.6/12.5  or 1.65.  The
ratio of the 20.6 \micron~size to the 12.5 \micron~size of the standard
star PSFs  observed  with \iras~16342$-$3814 is $\sim$1.6.
The ratio of sizes for  \iras~16342$-$3814 is a slightly smaller value
of 1.3 which we expect because \iras~16342$-$3814 is not a point source
and is better  resolved at 12.5 \micron~than at 20.6 \micron.

In five of the O-rich core/ellipticals,  the 9.8 \micron~image 
appears larger than the images at neighboring wavelengths 
(8.8, 11.7,  or 12.5 \micron).  
The source sizes for
\iras~07399$-$1435, \iras~16342$-$3814 and \iras~17347$-$3139
are notably  larger at 9.8 \micron~than at  neighboring
wavelengths.
For \iras~17150$-$3224 and 
\iras~19374+2359,
we find that the source is marginally larger at 9.8 \micron~than at 
12.5 \micron. In addition, 
\iras~18276$-$1431 (OH 17.7$-$2.0), which is also
an O-rich source,   is 
marginally resolved at 9.8 \micron~but unresolved at 12.5 \micron.
All six of these O-rich sources have a silicate absorption 
feature at 9.8 \micron~(see Table 1 for spectral reference and Table 3
for flux values).  Thus, the fact that the intensity at 9.8 \micron~
is more extended that the intensity at neighboring wavelengths appears
to  be correlated with the presence of a 9.8 \micron~silicate absorption
feature observed in these sources.

\iras~06176$-$1036 (Red Rectangle), which has 
both C-rich and O-rich dust components (cf. Table 1),
has an almost reversed appearance
compared to these O-rich sources.  It is smaller at 10 \micron~than at
its neighboring wavelengths, 8.2 and 11.3 \micron~as has also been found in
previous studies (\cite{hora96};
\cite{bregman93};\cite{sloan93}).  
The PAH emission features at 8.2 and 11.3 \micron~ (cf. Table 1) appear 
to be brighter in the low level elliptical nebula than
the 10\micron~ continuum dust emission which samples the non-PAH dust
grains and may be
sampling the O-rich dust component (Waters et al. 1998).

\subsubsection{Toroidals}

We classify
three objects as toroidals:
\iras~17436+5003, \iras~18184$-$1623 
and \iras~22223+4327 (Figure \ref{figmidirimag}). 
The toroidal sources  show evidence of a central torus  of dust.
\iras~17436+5003  (HD 161796) has  resolved  
inner contours  that are elongated  east-west this is  
the same direction
that \cite{skinner94} resolved two peaks  of a toroidal structure
at 12.5 \micron~in this object.  The cross cut of \iras~17436+5003
(Fig. \ref{figslice}) shows a  plateau-like, flattened
top which is expected when two limb brightened peaks are just beyond 
the resolution limit.  \iras~22223+4327 is similar
in morphology to  \iras~17436+5003 and has a plateau-like cross-cut similar to 
that of \iras~17436+5003.  Although we do not have other evidence
that this inner region will resolve into two limb-brightened peaks,
similarities between \iras~22223+4327 and the other toroidal sources found
in the literature  (discussed below) strongly suggest that it is toroidal.

\iras~18184$-$1623 shows 
a remarkable dust shell  which is clearly detached from
the central star.  Both 8.8 and 12.5$\mu$m
images show the central star surrounded by a partially complete
ring structure that has two peaks, almost symmetric around the star.
We interpret these two peaks as limb-brightened peaks of an
equatorial density enhancement,  i.e. a torus of dust; hence
we classify this object as a toroidal source.
If one were to  convolve \iras~18184$-$1623  with a PSF that had a FWHM 
equal to the separation of its two peaks,  one would find a source structure
similar to that in  \iras~17436+5003  and \iras~22223+4327. 
The  20.6 $\mu$m image also shows this dust shell structure;
however the central star is too faint to appear at the center.
Comparing all three wavelengths in \iras~18184$-$1623, we find
that the parts of the dust shell perpendicular to the torus
change. At 8.8 \micron, the dust shell is filled in at the southern
edge and partially filled at the northern edge.  As the wavelength
increases, we find this southern edge disappearing while the
northern edge gains more prominence, particularly at 20 \micron.
This change in morphology could be due to different temperatures
of the dust in these regions, i.e. the southern edge is warmer
than the northern edge, or, there could be differences in grain
types. Mid-IR spectroscopy of this nebula would help to
answer  these questions.
\iras~18184$-$1623 has been classified as a luminous blue variable (LBV),
and the toroidal structure that we observe here has been observed on 
comparable size scales at other wavelengths, e.g. 
H$\alpha$ (\cite{hutsemekers94}; \cite{nota96}),
near-IR and mid-IR (\cite{robberto98}).

\section{Discussion}

Mid-IR images of PPNe show us the size and morphology of the inner
regions of the dust shells in these sources.
If we enlarge the sample to include other published mid-IR imaging
studies, we can add eight additional
sources, seven of which are resolved (see Table 4B).
Of the 73 PPN candidates for which there exist mid-IR images,  
only 24 (33\%) are resolved.  Interestingly, two of the
unresolved sources, \iras~10158$-$2844 (HR 4049) and
\iras~15465+2818 (R CrB), have  extremely large  and cool (T$\sim$ 30 K) 
dust shells
ejected by an ancient mass loss episode (\cite{gillet86}, \cite{waters89}); 
the mid-IR images do not show these large dust shells, but rather
reveal  very recently ejected, compact dust shells.
Of the 24 resolved sources, 7 are
marginally resolved (unknown morphology), 10 have core/elliptical
extended structure, and 7 have toroidal extended structures.
The resolved sources are almost evenly divided between C-rich (13)
and O-rich (11) chemistry.  The core/elliptical sources seem
to be mostly O-rich (7 out of 10) and the toroidal sources
seem to be mostly C-rich  (5/7).  However, with such small numbers
(and presumably some selection effects) it is
hard to gauge the significance of these trends with chemistry.

There are three pieces of evidence suggesting that core/ellipticals
have denser dust shells than toroidals.
Firstly, comparison of the optical morphologies between toroidals  and
core/ellipticals shows significant differences (Table 4A and 4B).
In toroidal sources, the central star is very prominent in the optical
image and is  usually surrounded by nebulosity  that is faint in
comparison to the star. 
In contrast, the central star is rarely seen in the core/elliptical
sources and almost all of the  optical light is  reflection from
the nebula.  
The  morphologies of the optical  reflection nebulae   vary among the 
toroidals, but  almost all the core/ellipticals  display optical 
bipolar nebulae.
The optical bipolar reflection nebulae  found for core/ellipticals
extend in the same direction as the mid-IR elliptical nebula; however, the
optical bipolar nebulae are larger than the mid-IR elliptical nebulae. 
The difference in  optical morphology between toroidals and
core/ellipticals  suggests that toroidals
are transparent in the optical whereas  core/ellipticals are opaque in
the optical.

Secondly, the  core/ellipticals are typically O-rich   and
have deep silicate absorption
features  at 9.8 \micron~in the mid-IR spectra  which indicate that
they are optically thick in the mid--IR.  Our images of the core/ellipticals
show that the dust shells are compact  and that they appear
larger at 9.8 \micron~than at neighboring wavelengths (8 and 12.5 \micron).
In order to have an optically thick source which is compact, the dust shell
must be very dense.  The images together with the mid-IR spectra paint a
picture of dense ``dust photospheres'' that have decreasing  temperatures  
with increasing radial distance from the star.   At the more optically
thick wavelength of 9.8 \micron, we have a shallower view into the ``dust
photosphere'' and the dust shell appears both larger and fainter because
the dust shell is physically larger and cooler at this larger radial
distance from the star.  
At neighboring wavelengths, where the optical depth is lower,
we have a deeper view into the ``dust
photosphere'' and the dust shell appears smaller  and brighter 
because the dust shell is physically smaller and hotter at this
smaller radial distance from the star.      

Thirdly,  radiative transfer models of the star  and dust shell systems 
reveal marked differences in the densities of toroidals and core/ellipticals.
\cite{meixner97} model \iras~07134+1005, which is a toroidal,  C-rich PPN,
and find the dust torus to have low optical depth in the mid-IR 
($\tau_{9.7\mu m}\sim 0.03$) 
 and to have an inclination angle of $\sim 45^\circ$.  
Thus the fact that we see the central star in 
\iras~07134+1005 is not simply due to the viewing angle;  
we see the central star
because the dust density for \iras~07134+1005 is low.
On the other hand, \cite{skinner97} model the Egg Nebula (AFGL 2688),
which is a core/elliptical, C-rich PPN, and find the dusty torus  
to be optically thick not only in the optical, which has always been
presumed based on the obscuring dust lane of the bipolar reflection
nebula, but also in the mid-IR and longer wavelengths.  Both the high optical
depth at 10 \micron~($\tau_{10\mu m} \sim 2.4$) and
the compact size of the
dust shell indicate a high density for the inner regions of the Egg
Nebula.   These  model calculations corroborate the empirical evidence,
outlined above,
that core/ellipticals  have very dense cores which make them
optically thick at mid-IR wavelengths.
 The calculations also show
that toroidals are quite different in character to core/ellipticals,
i.e.  the difference is not just a viewing angle effect.

Why do core/ellipticals have higher density dust shells
than toroidals?  We suggest that core/ellipticals have experienced
overall higher mass loss rates compared to toroidals.
For example, the Egg Nebula, a core/elliptical, 
experienced a mass loss rate of $4\times 10^{-3}$\mloss~just
before it departed from the AGB (\cite{skinner97}); whereas  \iras~07134+1005,
a toroidal, experienced a significantly lower mass loss rate of 
$6\times 10^{-5}$
\mloss~(\cite{meixner97}).  Interestingly, both the Egg Nebula and
\iras~07134+1005 experienced
dramatic increases in mass loss rates by factors of $\sim$40 and 10,
respectively, prior to leaving the AGB (\cite{skinner97};\cite{meixner97}).
Hence, both types of sources, core/ellipticals and toroidals, have
experienced superwinds.   However, the important factor in creating 
a core/elliptical  as opposed to a toroidal source, is that the superwind
mass loss rate is very high; so high that it creates a dense, dusty torus
which is optically thick at mid-IR wavelengths and which obscures the 
central star at optical wavelengths.
What causes the higher mass loss rates in core/ellipticals
is not clear.  It may be that core/ellipticals have higher mass progenitor
stars  which are thought to experience higher mass loss rates at these
later stages.  

Future avenues of research with these mid-IR images are both theoretical
and observational.
The mid-IR images,  both resolved and unresolved,  are useful constraints
for modeling the circumstellar dust shells.  \cite{meixner97}
found that unresolved images provided upper limits on the inner radii of
the PPNe dust shells  and marginally resolved  images  provide a precise
constraint on the inner  radii.  Extended images  reveal not only the inner
radii, but also the geometry and density distribution
of the dust shells.  We plan to model
these sources, using additional information, 
to derive the physical parameters of these dust shells.  

Higher angular resolution would greatly increase the
number of resolved  sources.  What currently limits our angular
resolution at mid-IR wavelengths  is aperture size; i.e.
we are primarily diffraction limited.  With the new 8-10 m
aperture telescopes in operation and coming into operation in the
near future,   angular resolution will improve by a factor of two
making it  worthwhile to reimage all these sources.  

\section{Conclusions}

Of the 66 PPN candidates imaged at mid-IR wavelengths (8$-$25 \micron), 
17 were resolved, 48 were unresolved,
and 1 was not detected.  We present fluxes, sizes and morphologies
for the 17 resolved PPN candidates  and fluxes, and upper limits on sizes for
the 48 unresolved sources. These data will be useful for modeling the
circumstellar dust shells of these sources.  The 17 resolved sources
have brighter and  cooler (peaking at 25\micron) \iras~fluxes
than the unresolved sources which suggests that the resolved sources 
are closer or more luminous and  cooler (T $\sim$150 K) than
unresolved sources.  For the 11 extended sources,
two types of morphology are evident:  core/elliptical,
e.g. \iras~16342$-$3814,   
and   toroidal, e.g.  \iras~18184$-$1623.   We argue that this difference
in morphology corresponds to a difference in dust shell density:  
core/ellipticals  have significantly denser dust shells than toroidals.
The denser dust shells of core/ellipticals suggests that 
they originate from stars with higher
mass rates  than do the  toroidals.

 Even though we show
only a sub portion of the mid-IR images in  this paper, we  make all the mid-IR
images for all the sources and their PSFs available in FITS format 
 on the World Wide Web's NCSA Astronomy
Digital Image Library 
(ADIL, at URL: http://imagelib.ncsa.uiuc.edu/document/98.MM.02).

\acknowledgments

MIRAC2 has received support from
    NASA and NSF.  W. Hoffmann was visiting
    astronomer at the Institute for Astronomy, U. of Hawaii during
    part of this work.
Meixner and Ueta were  partially supported by the NSF (AST95-21605 and 
AST97-33697).  Hrivnak was partially supported by the
NSF (AST93-15107) and NASA (NAG 5-1223).  The Berkcam observations were partially
supported  by the Institute of Geophysics and Planetary Physics at
Lawrence Livermore National Laboratory.  
This research made use of the SIMBAD database, operated at CDS, Strasbourg, France, 
for the \iras~fluxes,
and many of the visual magnitudes and for a literature search of these
sources. We thank the helpful staff at the IRTF and UKIRT for their
assistance.  We are grateful to the  Hughes Aircraft Corporation, 
who supplied the mid-IR array used in Berkcam.  

\appendix

\section{Comments  on Objects in Figure  3}

In this appendix,
we make comments about each source shown in Figure 3.  The comments
are listed in the order that the source appears in Figure 3.

\iras~16559$-$2957 is unresolved and shows nice round PSFs.

\iras~22480$+$6002 is an  unresolved source with a ``cross pattern'' that 
is created by the IRTF astigmatism. 
Only the source and nearby K star are affected by IRTF astigmatism,
and therefore, it is difficult to recognize that the lowest contour 
structure is an artifact of IRTF astigmatism when one compares images 
of the source
and $\rm \beta Peg$, which is unaffected by the astigmatism because
of its lower declination.

\iras~02229$+$6208 is  marginally resolved at all five wavelengths.

\iras~06176$-$1036 (Red Rectangle) has
an unresolved core at all three
wavelengths, but is extended at 8.2 \micron, and 11.3 \micron~at 
the lower contour levels. Extension at 10 \micron~is also
seen at contour levels less than 10\%, the lowest contour level shown
here.

\iras~07399$-$1435 (OH 231.8+4.2) is extended at all five wavelengths, and
is  more extended at 9.8 \micron~than at 8.8 or 11.7 \micron.  The
general increase of size with increasing wavelength is a diffraction
effect (see section 3.5.1). 

\iras~16342$-$3814 is extended at all four wavelengths. Its
size increases from 12.5 \micron~to 20.6 \micron~as expected
from diffraction effects but it is larger at 9.8 \micron~than
12.5 \micron (see section 3.5.1).

\iras~17028$-$1004 (M 2$-$9) is unresolved at 8.1 and 8.5 \micron, 
marginally resolved at 9.7 \micron, and
extended at 12.6 and 12.8 \micron. 

\iras~17150$-$3224 is extended at all three wavelengths and is slightly
more extended at 9.8 \micron~than 8.1 or 12.5 \micron.
The 9.8 \micron~image shows the shape and PA
of the elliptical nebula best because the 8.1 and 12.5 \micron~images
are affected by artifacts seen in the PSFs.

\iras~17347$-$3139 is extended at both 9.8 and 12.5 \micron~but unresolved
at 20.6 \micron.  It is bigger at 9.8 \micron~than at 12.5 \micron
(see section 3.5.1).

\iras~17436$+$5003 (HD 161796) is extended at both wavelengths.
The inner elliptical contour extends in the same direction as the
two peaks resolved at 12.5 \micron~by \cite{skinner94}. 

\iras~17441$-$2411 is marginally extended at both wavelengths. 
The apparent ellipticity
and PA of its extended nebula is a little suspect because we
observe the some ellipticity with the same PA in the PSF.

\iras~18184$-$1302 (MWC 922) is marginally extended. 

\iras~18184$-$1623 (HD 168625) is extended at all three
wavelengths.  The central star, seen at the center, in the
8.8 and 12.5 \micron~images (but not the 20.6 \micron~image)
serves as a PSF reference.  The two peaks in the dust shell
are interpreted as limb-brightened peaks of an equatorial density
enhancement.  The images have been  smoothed by a Gaussian
($\sigma = 3$). The contour levels in this image start at 20\% and the
interval is 20\%.

\iras~18276$-$1431 (OH 17.7$-$2.3) is  marginally extended at 9.8 \micron~and 
unresolved at 12.5 \micron.

\iras~19244$+$1115 is extended at both wavelengths and 
exhibits an elliptical nebula.

\iras~19374$+$2359 is larger at 9.8 \micron~that the marginally extended
structures seen at  12.5 and 20.6 \micron.  

\iras~20000$+$3239 is marginally resolved at both 8.1 and 
12.5 \micron~(observed at 12.5 \micron~on two different date,
1995 June 3 and 6).  The core extension seen at 12.5 \micron~image 
taken in the June 3 run is suspect since $\rm \beta Peg$ observed
after the source (right) shows similar East-West elongation
and since the source image taken in the June 6 run does not 
show such extension.

\iras~22223$+$4327 is extended both at 12.5 and 18.0 \micron.
On both source images, structures suggested by lowest contours 
are not very reliable due to low signal-to-noise ratio.
North-South elongation of the inner-most core at 12.5 \micron~is
suspect since the PSF (middle frame) shows similar extension. 

\iras~23304$+$6147 is observed at 12.5 \micron~on two different dates, 1995 
June 5 and 6, and is extended, but is diffraction limited at 20.6 \micron.
The ``cross'' pattern seen at 10\% level of 12.5 \micron~source images
is certainly not real (IRTF astigmatism).

\newpage
\figcaption{\label{cartoon}   A schematic of our working model
for proto-planetary nebulae dust shells.  The axially symmetric
inner regions are created in a superwind mass loss phase.
The spherically symmetric outer regions are created in the AGB mass
loss phase.  The  shaded region of this schematic shows where the mid-IR
emission arises from the dust shell.   }

\figcaption{\label{figslice}  Representative examples of cross-cuts of 
the sources and PSF standards from each morphological class: 
(a) \iras~02229+6208 (marginally extended, unknown morphology), 
(b) \iras~16342$-$3814 (core/elliptical), and 
(c) \iras~17436+5003 (toroidal).
The observed wavelength and identity of
corresponding PSF are also indicated in the figure.  
Solid (dashed) lines show normalized intensity levels of the source 
(PSF) along the East-West
axis.  Notice the difference in emission levels around the mid-intensity 
region ($\sim 50\%$ level); 
(a) \iras~02229+6208 barely shows extended emission, 
(b) \iras~16342$-$3814 has a broad pedestal rising to a concentrated core
of emission,  and
(c) \iras~17436+5003 shows a plateau-like emission.
}

\figcaption{Mid-IR images of  PPN candidates and  their standard star PSFs.
Two unresolved PPN  candidates appear first followed by the resolved PPN
candidates which are shown in order of increasing RA. 
In each panel, the source appears in the left column, the standard star PSF
appears in the middle and right columns. When two PSFs are presented, the 
first (second) standard star was observed before (after) the source.  
The object name  appears in the top left corner of each image;
source names are  abbreviated to the RA portion of the \iras~name.
Each row of the panel
corresponds to a different mid-IR wavelength which appears in the bottom 
right corner of the image as a number in units of \micron.
The lowest contour levels are 10\% 
and the intervals between contour levels
are 20\% of the peaks of the images.  The peak values appear in
the bottom left corner of the image as a number in units of Jy arcsec$^{-2}$.
The coordinate axes show the relative displacement from the peak location
in arcseconds.   All the images presented, with the
exception of \iras~18184-1623, have the same
arcsecond/mm scale on the page for easy comparison.
Each row of the figures  corresponds to a particular wavelength.  
Important comments on each of these figures are listed in appendix A. 
\label{figmidirimag} }


\begin{thebibliography}{DUM}
 
\bibitem[Arens et al. (1987)]{arens87}
Arens, J. F., Jernigan, J. G., Peck, M. C., Dobson, C. A.,
Kilk, E., Lacy, J., \& Gaalema, S., 
1987, \ao,  26, 3846

\bibitem[Ball et al. (1992)]{ball92}
Ball, R., Arens, J. F., Jernigan, J. G., Keto, E., \& Meixner, M.
1992, \apj, 389,616

\bibitem[Bregman et al. (1993)]{bregman93}
Bregman, J.D., Rank, D., Temi, P., Hudgins, D. \& Kay, L. 1993, 411, 794  

\bibitem[Buss et al. (1990)]{buss90}
Buss, R. H., Jr. Cohen, M., Tielens, A. G. G. M., Werner, M. W., 
Bregman, J. D., Witteborn, F. C., Rank, D., \& Sandford, S. A. 
1990, \apj, 365, L23

\bibitem[Cohen \& Davies (1995)]{cohen95}
Cohen, M. \& Davies J. K. 
1995, \mnras, 276, 715

\bibitem[Cohen et al. (1975)]{cohen75}
Cohen et al.  
1975, \apj, 196, 179


\bibitem[Dayal (1997)]{dayal97}
Dayal, A.  
1997, PhD. Thesis, University of Arizona

\bibitem[Dayal et al. (1998)]{dayal98}
Dayal, A., Hoffmann, W.F., Bieging, J.H., Hora, J.L., Deutsch, L.K., \&
Fazio, G.G. 1998, \apj, 492, 603

\bibitem[Gillett \& Soifer (1976)]{gillet76}
Gillett, F. C.  \&  Soifer,  B. T. 
1976, \apj, 207, 780

\bibitem[Gillett et al. (1986)]{gillet86}
Gillett, F. C., Backman, D. E., Beichman, C., \& Neugebauer, G. 
1986, \apj, 310, 842

\bibitem[Groenewegen et al. (1998)]{groenewegen98}
Groenewegen, M. A. T., Whitelock, P. A., Smith, C. H., \& Kerschbaum, F. 
1998, \mnras, 293, 18
 
\bibitem[Hawkins et al. (1995)]{hawkins95}
Hawkins, G.W., Skinner, C.J., Meixner, M.M., Jernigan, J.G., Arens, J.F.
Keto, E., \& Graham, J.  1995, ApJ, 452, 314

\bibitem[Hoffmann et al. (1998)]{hoffmann98}
Hoffmann, W. F., Hora, J. L., Fazio, G. G., 
Deutsch, L. K.  \&  Dayal, A. 1998,  in Infrared Astronomical
    Instrumentation, ed. A. M. Fowler, Proc. SPIE 3354, in press
 
\bibitem[Hora et al. (1996)]{hora96}
Hora, J. L., Deutsch, L. K., Hoffmann, W. F., \&  Fazio, G. G. 
1996, \aj, 112, 2064

\bibitem[Hrivnak (1995)]{hrivnak95}
Hrivnak, B. J. 
1995, \apj, 438, 341

\bibitem[Hrivnak (1998)]{hrivnak98a}
Hrivnak, B. J. 1998, unpublished data

\bibitem[Hrivnak \& Kwok (1991)]{hrivnak91}
Hrivnak, B. J. \& Kwok, S. 
1991, \apj, 368, 564

\bibitem[Hrivnak, Kwok, \& Boreiko (1985)]{hrivnak85}
Hrivnak, B. J., Kwok, S., \& Boreiko, R. T. 
1985, \apj, 294, 113

\bibitem[Hrivnak \& Kwok  (1998)]{hrivnak98b}
Hrivnak, B. J. \& Kwok, S.  1998, \apj, in press

\bibitem[Hrivnak, Kwok  \&  Volk (1988)]{hrivnak88}
Hrivnak, B. J., Kwok, S.,\&  Volk, K. M.  
1988, \apj, 331, 832

\bibitem[Hrivnak, Kwok  \&  Volk (1989)]{hrivnak89}
Hrivnak, B. J., Kwok, S.,\&  Volk, K. M.  
1989, \apj, 346, 265

\bibitem[Hrivnak et al. (1999)]{hrivnak98c}
Hrivnak, B.J., Langill, P.P., Su, K.Y.L. \& Kwok, S.  1999, \apj, in press

\bibitem[Hu et al. (1993a)]{hu93a}
Hu, J. Y., Slijkhuis, S., De Jong, T., \&  Jiang B. W. 
1993a, \aaps, 100, 413

\bibitem[Hu et al. (1993b)]{hu93b}
Hu, J. Y., Slijkhuis, S., Nguyen-Q-Rieu \& De Jong, T. 
1993b, \aap, 273, 185

\bibitem[Humphreys et al. (1997)]{humphreys97}
Humphreys, R.M., Smith, N., Davidson, K., Jones, T.J., Gehrz, R.D., \&
Mason, C.G. 1997, \aj, 114, 2778

\bibitem[Hu et al. (1994)]{hu94}
Hu, J. Y., Te Lintel Hekkert, P., Slijkhuis, S., Baas, F., 
Sahai, R. \& Wood, P. R. 
1993, \aaps, 103, 301

\bibitem[Hutsemekers et al. (1994)]{hutsemekers94}
Hutsemekers, D., van Drom, E., Gosset, E., \& Melnick, J. 1994, \aap, 290, 906

\bibitem[Iyengar \& Parthasarathy (1997)]{iyengar97}
Iyengar, K. V. K., \& Parthasarathy, M. 
1997, \aaps, 121, 45

\bibitem[Justtanont et al. (1992)]{justtanont92}
Justtanont, K., Barlow, M. J., Skinner, C. J., 
\& Tielens, A. G. G. M. 
1992, \apj, 392, L75

\bibitem[Kastner et al. (1992)]{kastner92}
Kastner, J.H., Weintraub, D.A., Zuckerman, B., Becklin, E.E., McLean, I.,
\& Gatley, I. 1992, \apj, 398, 552 

\bibitem[Keto et al. (1992)]{keto92}
Keto, E., Ball, R., Arens, J. F., Jernigan, J. G. \& Meixner, M.
1992, International Journal of Infrared and Millimeter Waves, vol. 13,
N11:1709 

\bibitem[Kwok, Hrivnak \& Geballe (1995)]{kwok95}
Kwok, S., Hrivnak, B. J.,  \&  Geballe, T. R.  
1995, \apj, 454, 493
 
\bibitem[Kwok (1993)]{kwok93}
Kwok, S. 
1993, \araa, 31, 63
 
\bibitem[Kwok et al. (1996)]{kwok96}
Kwok, S., Hrivnak, B. J., Zhang, C. Y.  \&  Langill, P. L.  
1996, \apj, 472, 287

\bibitem[Kwok, Su \& Hrivnak (1998)]{kwok98}
Kwok, S., Su, K.Y.L.,  \& Hrivnak, B.J., 1998, \apj, 501, L117

\bibitem[Likkel (1989)]{likkel89}
Likkel, L.  
1989, \apj, 344, 350

\bibitem[Loup et al. (1993)]{loup93}
Loup, C., Forveille, T., Omont, A. \& Paul, J. F. 
1993, \aaps, 99, 291

\bibitem[Manchado et al. (1989)]{manchado 89}
Manchado, A., Pottasch, S. R., Garcia-Lario, P., Esteban, C., \& Mampaso, A.
1989, \aap, 214, 139

\bibitem[Mauron et al. (1989)]{mauron89}
Mauron, N., Le Borgne J.-F., \& Picquette, M. 
1989, \aap, 218, 213

\bibitem[Meixner (1993a)]{meixner93a}
Meixner, M. 
1993, PhD Thesis,  UC Berkeley

\bibitem[Meixner (1993b)]{meixner93b}
Meixner, M., et al. 1993, \apj, 411, 266 


\bibitem[Meixner et al. (1997)]{meixner97}
Meixner, M., Skinner, C. J., Graham, J. R., Keto, E., Jernigan, J. G.,
\& Arens, J. F.  
1997, \apj, 482, 897

\bibitem[Ney et al. (1975)]{ney75}
Ney, E.P., Merriall, K.M., Becklin, E.E., Neugebauer, G., \& 
Wynn-Williams, C.G., 1975, \apj, 198, L129

\bibitem[Nota et al. (1996)]{nota96}
Nota, A., Pasquali, A., Clampin, M., Pollacco, D., Scuderi, S., \& Livio, M.
1996, \apj, 473, 946

\bibitem[Olnon \& Raimond (1986)]{olnon86}
Olnon, F. M., Raimond, E., \& IRAS Science Team  
1986, \aaps, 65 607

\bibitem[Omont et al. (1993)]{omont93}
Omont, A., Loup, C., Forveille, T., Te Lintel Hekkert, P., Habing, H.,\&
Sivagnanam, P. 
1993, \aap, 267, 515

\bibitem[Oudmaijer et al. (1992)]{oudmaijer92}
Oudmaijer, R.D., van der Veen, W. E. C. J., Waters, L. B. F. M., 
Trams, N. R., Waelkens, C., \& Engelsman, E. 
1992, \aaps, 96, 625

\bibitem[Robberto \& Herbst (1998)]{robberto98}
Robberto, M. \& Herbst, T. M.  
1998, \apj, 498, 400

\bibitem[Silva et al. (1993)]{silva93}
Silva, A. M., Azcarate, I. N., Poppel, W. G. L. \& Likkel, L.  
1993, \aap, 275, 510

\bibitem[Skinner et al. (1994)]{skinner94}
Skinner, C. J., Meixner, M., Hawkins, G. W., Keto, E., Jernigan, J. G. 
\& Arens, J. F. 
1994, \apj, 423, 135

 
\bibitem[Skinner et al. (1997)]{skinner97}
Skinner, C.J., Meixner, M., Barlow, M.J., Collison, A.J.,
Justtanont, K., Blanco, P., Pi\~{n}a, R., Ball, J.R., Keto, E.,
Arens, J.F., \& Jernigan, J.G. 1997, \aap, 328, 290
 
\bibitem[Sloan, Grasdalen \& Levan (1993)]{sloan93}
Sloan, G.C., Grasdalen, G.L. \& Levan, P.D. 1993, \apj, 409, 412

\bibitem[Trammell, Goodrich \& Dinerstein (1995)]{trammell95} 
Trammell S.R., Goodrich, R.W., \& Dinerstein, H.L. 1995, \apj, 453, 761

\bibitem[Te Lintel Hekkert (1991)]{lintel91}
Te Lintel Hekkert,  P. 1991, \aap, 248, 209

\bibitem[Ueta, Meixner \& Bobrowsky (1998)]{meixner98}
Ueta, T., Meixner, M., \& Bobrowsky, M.  1998, in preparation

\bibitem[van der Veen, Habing \& Geballe (1989)]{veen89}
van der Veen, W. E. C. J., Habing, H. J.  \&  Geballe, T. R.  
1989, \aap, 226, 108

\bibitem[Volk \& Cohen (1989)]{volk89a}
Volk, K., \& Cohen, M. 
1989, \aj, 98, 931

\bibitem[Volk \& Cohen (1990)]{volk90}
Volk, K., \& Cohen, M. 
1990, \aj, 100, 485

\bibitem[Volk \& Kwok (1989)]{volk89b}
Volk, K., \& Kwok, S. 
1989, \apj, 342, 345

\bibitem[Volk, K., Kwok, S. \& Langill, P. P. (1992)]{volk92}
Volk, K.,  Kwok, S., \&  Langill, P. P.  
1992,  \apj,  391,  285

\bibitem[Volk, Kwok, \& Woodsworth (1993)]{volk93}
Volk, K., Kwok, S., \& Woodsworth, A. W.  
1993,  \apj, 402, 292

\bibitem[Waters et al. (1989)]{waters89}
Waters, L. B. F. M., Lamers, H. J. G. L. M., Snow, T. P., 
Mathlener, E., Trams, N. R., van Hoof, P. A. M., Waelkens, C., 
Seab, C. G., \& Stanga, R.  
1989, \aap, 211, 208

\bibitem[Waters et al. (1998)]{waters98}
Waters, L.B.F.M., Waelkens, C., van Winckel, H., Molster, F.J., Tielens,
A.G.G.M., van Loon, J. TH., Morris, P.W., Cami, J., Bouwman, J., de
Koeter, A., de Jong, T., \& de Graaum, Th  1998, Nature, 391, 868

\bibitem[Westbrook et al. (1975)]{westbrook75}
Westbrook, W.E., Becklin, E.E., Merrill, K.M., Neugebauer, G., Schmidt, M.
Willner, S.P., \& Wynn-Williams, M. J. 1986, \apj, 304, 401

\bibitem[Young, Phillips \& Knapp (1993)]{young93}
Young, K., Phillips, T. G. \& Knapp, G. R. 
1993, \apjs, 86, 517

\bibitem[Zijlstra et al. (1989)]{zijlstra89}
Zijlstra, A. A., Te Lintel Hekkert, P., Pottasch, S. R., 
Caswell, J. L., Ratag, M., \& Habing, H. J.  
1989, \aap, 217, 157

\end{thebibliography}
\end{document}


\pagestyle{empty}

\def\ell{ .~.~.}
\begin{deluxetable}{lcrrrrrrccc}

\tablenum{1}
\tablewidth{0pt}
\tablecaption{Characteristics  of Proto-Planetary  Nebulae  in Mid-IR  Survey}

\tablehead{
\colhead{\iras~ID} & \colhead{C/O\tablenotemark{a}} & 
\colhead{Sp. Type} & 
\colhead{V} & 
\colhead{F$_{\mbox{12}}$ [Jy]} & 
\colhead{F$_{\mbox{25}}$ [Jy]} & 
\colhead{F$_{\mbox{60}}$ [Jy]} & 
\colhead{F$_{\mbox{100}}$ [Jy]} & 
\colhead{Mid-IR Sp.\tablenotemark{b}} &  \colhead{D[kpc]} &
\colhead{Ref.\tablenotemark{c}}}  
 
\startdata

00210+6221\tablenotemark{ii} & C? &  \ell & 16.5 &48.5 &51.9 &12.5 & \ell &1 &\ell & A,W,Z\nl
00470+6429  &\ell &\ell & 12.0 & 4.6 & 4.9 & 2.1 & 10.5 &\ell &\ell &\ell \nl
01144+6658\tablenotemark{i}  & C &\ell &\ell & 139.9 & 206.1 & 64.7 & 15.9 &1 & 4 & A,B \nl
01174+6110  & C &\ell & 18.3 & 4.1 & 16.9  & 33.9 & 30.7 &\ell &\ell & C,Z \nl
01304+6211  &O &\ell &\ell & 289.0 & 455.6 & 194.0 & 50.2 &1 & 1.5 & E \nl
02143+5852  & O & F & 13.8 & 5.9 & 18.1 & 5.4 & \ell &\ell &\ell & \ell \nl
02152+2822\tablenotemark{i}  & C &C &\ell & 120.5 & 112.2 & 29.6 & 6.9 &1 &2.1 & E,F\nl
02229+6208  & C & G8-K0 Ia & 12.1 & 67  & 204 & 31 & \ell &\ell &\ell & V \nl
04386+5722  &\ell & M & 12.7 & 18 &23 & 4.4 & \ell &\ell &\ell & J \nl
05089+0459  & C? & M & 14.3 & 7.4 & 21.9 & 11.1 & 3.8 &\ell &\ell &J \nl
05113+1347  &C & G8Ia & 12.4 & 3.8 & 15.3 & 5.5 & \ell & 2 &\ell &J,U \nl
05341+0852  &C & G2 0-Ia & 13.6 & 4.5 & 9.9 & 4.0 & \ell & 2 &\ell & V \nl
06176$-$1036  &C/O & B8V &8.8 &422 &456 &173 &66 & 1 &0.33 & I,AA \nl
06530$-$0213  &C/O & F0Iab & 14.0 & 6.1 & 27.4 & 15.1 & 4.1 &\ell & 1.3 & G,H,E \nl
07239$-$3325  &\ell &\ell &\ell & 1 & 2.4 & \ell & \ell & \ell&\ell &\ell \nl
07284$-$0940\tablenotemark{vi}  &O & F8Ibe &6.3 & 124 & 88.4 & 26.6 & 9.5 &1 &\ell &F\nl
07331+0021  &O & G5Iab & 8.8 & 15.3 & 68.1 & 18.5 &3.7 &\ell &\ell &F \nl
07399$-$1435\tablenotemark{i} & O & M6 & 9.5 & 19 & 226.3 & 548.3 & 294.0 & 1,3 &1.3 & E \nl
07430+1115  &C & G5 0-Ia &12.6 & 7.7 & 29.9 & 10.7 & 2.5 &\ell &\ell & J,V \nl
08005$-$2356  &C/O & F5e & 11.5 & 18  & 51.8 & 29.8 & 10.4 &\ell &\ell & J,K \nl
09032$-$3953  &\ell &\ell & 18.3 &20.4 & 125.9 & 92.3 & 28.7 &1 &\ell & G \nl
09371+1212  &O & K7II & 11.0 & 0.3L & 4.6 & 70.7 & 28.2 &\ell &\ell & E,L \nl
10131+3049  & C & M &9.4 & 3319 & 1219 & 273.6 & 86.1 & 1 &0.19 & E \nl
10158$-$2844  &C &B9.5Ib-II &5.3 &48 &9.6 &1.8 & \ell &1, 4 &3.5 &\ell \nl
11472$-$0800  &\ell &\ell &12 & 11.4 & 14.2 & 1.8 & \ell &\ell &\ell &\ell \nl
15465+2818  &C &F0-F8pe & 5.9 & 38.9 &17.1 &3.94 &2.0 &1,4 & 1.3 &\ell \nl
16342$-$3814\tablenotemark{i}  & O & M & 13.7 & 16.2 & 199.8 & 290.2 & 139.4 &1 & 2.2 & M \nl
16559$-$2957  &O & F5Ie & 13.2 & 9.2 & 32.4 & 16.4 & \ell &\ell &\ell & G,K \nl
17028$-$1004\tablenotemark{ii}  & O & B1 & 14.7 & 50.5 & 110.2 & 123.6 & 75.8 &5,6 &1.6 &\ell \nl
17088$-$4221  & O & A0 &\ell & 42.7 & 128.3 & 106.8 & 36.9 &1 & 2.4 & M \nl
17106$-$3046  & O &\ell & 15.8 & 4.0 & 62.4 & 51.2 & 17.3 &\ell &\ell & N,Z \nl
17150$-$3224  &O &G2 & 14.5 &58 & 322.3 & 268.3 & 82.4 & 1,2 & 2.42 & M,E,O \nl
17245$-$3951  &\ell &\ell & 15.5 & 3.4 & 44.7 & 38.2 & \ell &1 &\ell & Z \nl
17347$-$3139\tablenotemark{ii}   &O &O &\ell & 19 & 100.3 & 125.4 & 249.0 &\ell &\ell & P \nl
17436+5003  &O & F3Ib & 7.1 & 6.12 & 184 & 152 &48.7 & 1,7 &1.2 & Q \nl
17441$-$2411  & O &\ell & 15.6 & 42.8 & 191.1 & 106.5 & \ell &1 & 1.9 & R \nl
17534+2603  &O & F2Ibe  & 5.5 & 97.5 & 54.5 & 13.4 & 6.04 &1 &1 &\ell \nl
18071$-$1727  &O &\ell &\ell & 23.1 & 76.6 & 83.5 & \ell &1 &\ell &F, M \nl
18095+2704  &O & F3Ib & 10.4 & 45.1 & 125.7 & 27.8 & 5.6 &1 &1.9 & Y \nl
18184$-$1302\tablenotemark{iv}  &C &Be & 13.9 & 336.0 & 597.5 & 253.2 & 436.0 &1 &1.7 & \ell\nl
18184$-$1623\tablenotemark{v}   &C & O+ & 8.4 & 70 & 325 & 117 & 584 &5 & 1.2 &T \nl
18276$-$1431  &O & K5-B9 &\ell & 22.7 & 132.0 & 120.0 & 38.6 &1 &\ell & F \nl
19075+0921\tablenotemark{i}  &C &M &\ell & 133.0 & 163.7 & 54.9 & \ell &1 & 4 & A,F \nl
19192+0922\tablenotemark{i}  &O &M &\ell & 127.2 & 155.0 & 41.4 & 9.8 &1 &1.3 & E \nl
19244+1115\tablenotemark{iii}  &O & F8Ia & 8.5 & 1346.0 & 2314.0 & 717.7 & 185.8 &1 & 3.4 & E \nl
19343+2926  & O &B0.5IV & 11.7 & 17.5 & 59.8 & 118.4 & 68.0 &\ell  & 2.5 &\ell \nl
19374+2359  &O &\ell & 18.1 & 23.6 & 98.2 & 70.9 & \ell & 8 & 5.2 & M,Z \nl
19386+0155  &\ell &F & 11.2 & 17.4 & 47.4 & 18.6 & 3.8 &\ell & 5.3 & M \nl
19454+2920  &C &\ell &\ell &17.3 & 89.6 & 54.4 & 14.7 &1 &\ell & A,F \nl
19477+2401  &O &\ell & 22.2 & 11.2 & 54.9 & 27.1 & \ell & 2 &\ell & Z \nl
19480+2504  &C &\ell &\ell & 20.8 & 67.9 & 43.2 & \ell &1 &\ell & A,F \nl
20000+3239  &C & G8Ia & 13.2 & 15.0 & 71 & 30 & \ell & 2  &\ell & F,U \nl
20004+2955  & O & G7Ia & 8.9 & 31.7 & 37 & 4.7 & \ell & 1 &\ell & S \nl
20028+3910  &\ell &\ell & 17.8 & 41.8 & 210.8 & 143.1 & 46.5 &1 &\ell &F,Z \nl
20043+2653  & O &\ell &\ell & 17.9 & 42 & 20.3 & 7.5 &1 &\ell & D \nl
20077$-$0625  &O & M & 8.5 & 1255 & 1061 & 215.5 & 63.7 & 1 & 0.65 &E \nl
21537+6435  &\ell &\ell &\ell & 6.9 & 26.1 & 13.3 & \ell &\ell &\ell &\ell \nl
22036+5306\tablenotemark{c}  & O &\ell &\ell & 8.4 & 46.3 & 107 &50.7 &1 &\ell & C \nl
22142+5206\tablenotemark{d}  & O &\ell & 21.1 & 20.8 & 92.4 & 171.5 & 139.6 & 8 &\ell &\ell \nl
22223+4327  & C & G0Ia & 9.7 & 2.1 & 37.1 & 22.4 & 9.5 & 2 &\ell & F,U \nl
22480+6002  &O &K0:Ia & 10.4 & 141.9 & 175.2 & 39.8 & \ell &1 &\ell &\ell \nl
22574+6609  &C &\ell & 24 & 9.0 & 29.5 & 20.6 & 7.2 &\ell &\ell & X \nl
23166+1655\tablenotemark{i}  &C &M &\ell & 706.7 & 775.6 & 248.5 & 73.7 & 1 & 1.2 & B \nl
23304+6147  & C & G2Ia & 13.1 & 11.4 & 59.1 & 26.6 & \ell &\ell &\ell & F,U \nl
23321+6545  &C/O &\ell &\ell & 13.7 & 85.6 & 64 & 19.3 &1 &\ell & A,F,H \nl
\tablenotetext{a}{ C = carbon rich; O = oxygen rich; C/O = both carbon and oxygen rich }
\tablenotetext{b}{ References for Mid-IR spectra: 
1. Olnon \& Raimond (1986; \iras~LRS Atlas); 
2. Kwok, Hrivnak \& Geballe (1995); 
3. Gillet \& Soifer (1976); 
4. Buss et al. (1990); 
5. Volk \& Cohen (1989); 
6. Volk \& Cohen (1990); 
7. Justtanont et al. (1992); 
8. Volk et al. (1991)}
\tablenotetext{c}{References  for other information: 
SIMBAD for \iras~measurements of all sources and for V and spectral type for many sources; 
Oudmaijer et al. (1992) for V and spectral type of many sources; 
Hrivnak (1998) for V and spectral type of many sources; 
A. Volk, Kwok \& Woodsworth (1993); 
B. Groenewegen et al. (1998);  
C. Te Lintel Hekkert  (1991);  
D. Manchado et al.  (1989); 
E. Loup et al. (1993);  
F. Omont et al. (1993);  
G. Hu et al. (1993a);  
H. Likkel (1989);  
I. Cohen et al. (1975); 
J. Iyengar \& Parthasarathy (1997); 
K. Hu et al. (1994); 
L. Mauron, Le Borgne \& Picquette (1989);  
M. van der Veen, Habing \& Geballe (1989);  
N. Silva et al. (1993);  
O. Hu et al. (1993b);  
P. Zijlstra et al. (1989);  
Q. Skinner et al. (1994);  
R. Kwok et al. (1996); 
S. Hrivnak, Kwok \& Volk (1989);  
T. Robberto \& Herbst (1998); 
U. Hrivnak (1995); 
V. Hrivnak \& Kwok (1998); 
W. Hrivnak, Kwok, \& Boreiko (1985); 
X. Hrivnak \& Kwok (1991);
Y. Hrivnak, Kwok \& Volk (1988);
Z. Hrivnak et al. (1999);
AA. Waters et al. (1998)}
\tablenotetext{c}{HD235718 (a.k.a. SAO 34043), a V=9.5 magnitude B9
star  is NOT associated with this source.  See text.}
\tablenotetext{d}{This source is located 7\arcsec~north of \iras~coordinates.  
See text.}
\tablenotetext{}{}
\tablenotetext{i}{extreme AGB star}
\tablenotetext{ii}{very young planetary nebula}
\tablenotetext{iii}{supergiant}
\tablenotetext{iv}{probably pre-main sequence star}
\tablenotetext{v}{luminous blue variable (LBV)}
\tablenotetext{vi}{RV Tauri star}
\enddata
\end{deluxetable}

\begin{deluxetable}{lrrccc}

\tablenum{2}
\tablewidth{0pt}
\tablecaption{MIRAC2 and Berkcam Observing Parameters}

\tablehead{
\colhead{Identification} & 
\colhead{RA(2000)} & \colhead{DEC(2000)} & 
\colhead{Date, Tel\tablenotemark{a}, Cam\tablenotemark{b}} & \colhead{$\lambda$ ($\Delta\lambda$) [$\mu m$]} & 
\colhead{Standard}}
 
\startdata
IRAS 00210+6221 & 
00:23:51.1 & +62:38:07 & 11/7/95, I, M & 12.5 (10.0$\%$) & $\beta$And \nl

IRAS 00470+6429 & 
00:50:07.2 & +64:45:46 & 11/10/95, I, M & 10.6 (8.1-13.1) & $\beta$And \nl

IRAS 01144+6658 & 
01:17:51.6 & +67:13:52 & 11/10/95, I, M & 8.8 (10.0$\%$) & $\beta$And \nl
~~(AFGL 190) &  &  & 11/10/95, I, M & 12.5 (10.0$\%$) & $\beta$And \nl

IRAS 01174+6110 & 
01:20:44.3 & +61:26:21 & 11/7/95, I, M & 11.7 (10.0$\%$) & $\beta$And \nl

IRAS 01304+6211 & 
01:33:51.2 & +62:26:53 & 11/7/95, I, M & 9.8 (10.0$\%$) & $\beta$And \nl
~~(OH 127.8+0.0) \nl
~~(V* V669 Cas) \nl

IRAS 02143+5852 & 
02:17:57.9 & +59:05:51 & 11/22/95, U, M & 11.7 (10.0$\%$) & $\beta$Peg \nl

IRAS 02152+2822 &  
02:18:06.6 & +28:36:48  & 12/14/92, U, B & 11.3CVF (10\%) & $\beta$And\\

IRAS 02229+6208 & 
02:26:41.8 & +62:21:22 & 11/7/95, I, M & 8.8 (10.0$\%$) & $\beta$And \nl
 &  &  & 11/7/95, I, M & 9.8 (10.0$\%$) & $\beta$And \nl
 &  &  & 11/7/95, I, M & 11.7 (10.0$\%$) & $\beta$And \nl
 &  &  & 11/7/95, I, M & 12.5 (10.0$\%$) & $\beta$And \nl
 &  &  & 11/7/95, I, M & 20.6 (7.0$\%$) & $\alpha$Tau \nl
 
HD 20041 &  
03:15:48.0  &  +57:08:26.2 & 11/22/95,U, M & 11.7 (10.0\%) &  $\beta$Peg \nl

IRAS 04386+5722 & 
04:42:48.3 & +57:27:46.7 & 11/10/95, I, M & 12.5 (10.0$\%$) & $\alpha$Tau \nl

IRAS 05089+0459 & 
05:11:36.0 & +05:03:26 & 11/22/95, U, M & 11.7 (10.0$\%$) & $\alpha$Tau \nl

IRAS 05113+1347 & 
05:14:07.8 & +13:50:28 & 11/7/95, I, M & 11.7 (10.0$\%$) & $\alpha$Tau \nl

IRAS 05341+0852 & 
05:36:55.0 & +08:54:09 & 11/7/95, I, M & 8.8 (10.0$\%$) & $\alpha$Tau \nl
 &  &  & 11/7/95, I, M & 11.7 (10.0$\%$) & $\alpha$Tau \nl
 &  &  & 11/23/95, U, M & 11.7 (10.0$\%$) & $\alpha$Tau \nl

IRAS 06176$-$1036 & 
06:19:58.2 & $-$10:38:13.7 & 12/13/92, U, B &  8.2CVF (10\%) & $\alpha$CMa \nl 
~~(Red Rectangle) &  &  & 12/13/92, U, B  &  10.0CVF (10\%) & $\alpha$CMa \nl 
~~(HD 44179) &  &  & 12/9/92, U, B &  11.3CVF (10\%) &  $\alpha$CMa \nl
~~(AFGL 915) \nl

IRAS 06530$-$0213 & 
06:55:30.5 & $-$02:17:25 & 11/10/95, I, M & 12.5 (10.0\%) &  $\alpha$Tau  \nl

IRAS 07239$-$3325 & 
07:25:52.2 & $-$33:31:22 & 11/23/95, U, M & 10.6 (8.1-13.1) & $\alpha$CMi \nl

IRAS 07284$-$0940 & 
07:30:47.4 & $-$09:46:37 & 11/23/95, U, M & 9.8 (10.0$\%$) & $\alpha$CMi \nl
~~(HD 59693) &  &  &  &  &  \nl
~~(V* U Mon) &  &  &  &  &  \nl
 
IRAS 07331+0021 & 
07:35:41.1 & +00:14:59 & 11/23/95, U, M & 8.8 (10.0$\%$) &  $\alpha$CMa, $\alpha$Tau \nl
~~(V* AI CMi) &  &  & 11/23/95, U, M & 20.6 (7.0$\%$) & $\alpha$CMi \nl

IRAS 07399$-$1435 & 
07:42:16.8 & $-$14:42:52 & 11/6/95, I, M & 8.8 (10.0$\%$) & $\alpha$CMa \nl
~~(OH 231.8+4.2) & & & 11/6/95, I, M & 9.8 (10.0$\%$) &$\alpha$CMa \nl
 & & & 11/6/95, I, M & 11.7 (10.0\%) &$\alpha$CMa \nl
 & & & 11/6/95, I, M & 18.0 (10.0\%) &$\alpha$CMa \nl
 & & & 11/6/95, I, M & 20.6 (7.0\%) &$\alpha$CMa \nl

IRAS 07430+1115 & 
07:45:51.4 & +11:08:16 & 11/23/95, U, M & 8.8 (10.0$\%$) & $\alpha$Tau \nl
 &  &  & 11/7/95, I, M & 11.7 (10.0$\%$) & $\alpha$Tau \nl

IRAS 08005$-$2356 & 
08:02:41.9 & $-$24:04:48 & 11/22/95, U, M & 9.8 (10.0$\%$) & $\alpha$CMi, $\beta$Peg \nl
 
IRAS 09032$-$3953 & 
09:05:07.2 & $-$40:05:11 & 11/22/95, U, M & 8.8 (10.0$\%$) & $\alpha$CMi \nl
 
IRAS 09371+1212 & 
09:39:53.6 & +11:58:54 & 11/23/95, U, M & 3.8 (16.0$\%$) & $\alpha$CMa \nl
(Frosty Leo) &  &  & 11/23/95, U, M & 10.6 (8.1-13.1) &  $\alpha$CMi \nl

\tablebreak

IRAS 10131+3049 & 
10:16:02.0  & +30:34:19 & 6/9/92, U,B  &  11.5 (10\%) &  $\mu$UMa\nl
~~(CIT 6) \nl
~~(IRC +30219) \nl

IRAS 10158$-$2844 &  
10:18:07.6 & $-$28:59:32 &  6/8/92, U, B & 8.5CVF (10\%) & $\alpha$Boo, BS4104 \nl
~~(HR 4049) &  & &  6/7/92, U, B  &  9.7CVF (10\%)  &  $\alpha$Boo, BS4104 \nl
~~(HD 89353)\nl

IRAS 11472$-$0800 & 
11:49:48.4 & $-$08:17:21 & 11/24/95, U, M & 9.8 (10.0$\%$) &  $\delta$Vir, $\mu$UMa \nl
 
IRAS 15465+2818 & 
15:48:34.3 & +28:09:25 &  5/15/91, I, B & 12.5CVF (10\%)  &  $\alpha$Boo\nl
~~(R CrB) \nl

IRAS 16342$-$3814 & 
16:37:40.0 & $-$38:20:14 & 6/3/95, I, M & 9.8 (10.0$\%$) & $\sigma$Lib \nl
 &  &  & 6/3/95, I, M & 12.5 (10.0$\%$) &  $\alpha$Boo, $\sigma$Lib\nl
 &  &  & 6/3/95, I, M & 17.8 (2.6$\%$) & $\eta$Sgr \nl
 &  &  & 6/3/95, I, M & 20.6 (7.0$\%$) &  $\sigma$Lib, $\eta$Sgr \nl

IRAS 16559$-$2957 & 
16:59:08.1 & $-$30:01:41 & 6/3/95, I, M & 12.5 (10.0$\%$) & $\eta$Sgr \nl
 
IRAS 17028$-$1004 & 
17:05:37.9  &  $-$10:08:27 & 5/16/91, I, B & 8.1CVF (1.3\%) & $\alpha$Her  \nl
~~(M 2$-$9) &  &  & 5/16/91, I, B &  8.5CVF (10\%)  & $\alpha$Her \nl
 &  &  & 5/16/91, I, B &  9.7CVF (1.3\%) & $\alpha$Her  \nl
 &  &  & 5/16/91, I, B &  12.6CVF (1.3\%) & $\alpha$Her  \nl
 &  &  & 5/16/91, I, B &  12.8CVF (1.3\%) & $\alpha$Her  \nl

IRAS 17088$-$4221 & 
17:12:21.5 & $-$42:25:08 & 6/3/95, I, M & 12.5 (10.0$\%$) & $\eta$Sgr \nl

IRAS 17106$-$3046 & 
17:13:51.6 & $-$30:49:42 & 6/7/95, I, M & 18.0 (10.0$\%$) & $\eta$Sgr \nl

IRAS 17150$-$3224 & 
17:18:20.0 & $-$32:27:23 & 6/5/95, I, M & 8.1CVF (1.8$\%$) &  $\eta$Sgr, $\alpha$Sco \nl
 &  &  & 6/5/95, I, M & 9.8 (10.0$\%$) & $\alpha$Sco \nl
 &  &  & 6/3/95, I, M & 12.5 (10.0$\%$) & $\eta$Sgr \nl

IRAS 17245$-$3951 & 
17:28:04.7 & $-$39:53:44 & 6/3/95, I, M & 12.5 (10.0$\%$) & $\eta$Sgr \nl

IRAS 17347$-$3139 & 
17:38:00.6 & $-$31:40:54 & 6/5/95, I, M & 9.8 (10.0$\%$) &  $\eta$Sgr, $\alpha$Sco \nl
 &  &  & 6/5/95, I, M & 12.5 (10.0$\%$) &  $\eta$Sgr, $\alpha$Sco \nl
 &  &  & 6/5/95, I, M & 20.6 (7.0$\%$) &  $\eta$Sgr, $\alpha$Sco \nl

IRAS 17436+5003 & 
17:44:55.43 & +50:02:39.9 & 6/7/95, I, M & 18.0 (10.0$\%$) & $\gamma$Dra \nl
~~(HD 161796) &  &  & 6/7/95, I, M & 20.6 (7.0$\%$) & $\gamma$Dra \nl
~~(V* V814 Her) &  &  &  &  &  \nl

IRAS 17441$-$2411 & 
17:47:13.0 & $-$24:12:54 & 6/3/95, I, M & 11.3CVF (1.8$\%$) &  $\eta$Sgr, $\alpha$Her \nl
 &  &  & 6/3/95, I, M & 12.5 (10.0$\%$) & $\eta$Sgr \nl

IRAS 17534+2603 & 
17:55:25.1 & +26:03:00 &  5/15/91, I, B & 12.5CVF (10\%) &  $\alpha$Her\nl
~~(89 Her) \nl
~~(HD 163506)\nl

IRAS 18071$-$1727 & 
18:10:05.9 & $-$17:26:35 & 6/6/95, I, M & 9.8 (10.0$\%$) & $\eta$Sgr \nl
 &  &  & 6/6/95, I, M & 12.5 (10.0$\%$) & $\eta$Sgr \nl

IRAS 18095+2704 &   
18:11:30.6 & +27:05:16 &  6/8/92, U, B &  8.5CVF (10\%) & $\alpha$Lyr\nl
~~(OH 53.8+20.2) & & & & 11.0 (10\%) & $\alpha$Lyr \nl 

IRAS 18184$-$1302 &  
18:21:16 & $-$13:01:27 & 6/9/92, U, B &  8.2CVF (10\%) &  $\alpha$Lyr \nl
~~(MWC 922) & & & &  12.2 (10\%) &  $\alpha$Lyr  \nl

IRAS 18184$-$1623 & 
18:21:19.46 & $-$16:22:26.0 & 6/4/95, I, M & 8.8 (10.0$\%$) & $\sigma$Lib \nl
~~(HD 168625) &  &  & 6/5/95, I, M & 12.5 (10.0$\%$) & $\eta$Sgr \nl
~~(V* V4030 Sgr) &  &  & 6/3/95, I, M & 12.5 (10.0$\%$) & $\beta$Peg \nl
 &  &  & 6/4/95, I, M & 20.6 (7.0$\%$) & $\alpha$Her \nl

\tablebreak

IRAS 18276$-$1431 & 
18:30:30.37 & $-$14:28:53.2 & 6/6/95, I, M & 9.8 (10.0$\%$) &  $\alpha$Lyr, $\alpha$Sct \nl
~~(OH 17.7$-$2.0) &  &  & 6/6/95, I, M & 12.5 (10.0$\%$) &  $\alpha$Lyr, $\alpha$Sct \nl
 
IRAS 19075+0921 & 
19:09:57.1 & +09:26:52 & 6/9/92, U, B &  8.2CVF (10\%) & $\alpha$Lyr \nl
~~(AFGL 2333) & & & 6/9/92, U, B & 12.2CVF (10\%) &  $\alpha$Lyr \nl

IRAS 19192+0922 & 
19:21:36.5 & +09:27:56 &  6/8/92, U, B &  12.5CVF (10\%) &  $\gamma$Aql\nl
~~(OH 44.8$-$2.3)\nl

IRAS 19244+1115 & 
19:26:48.30 & +11:21:08.7 & 6/4/95, I, M & 9.8+20$\%$ (10.0$\%$) &  $\alpha$Aql \nl
~~(IRC +10420) &  &  & 6/4/95, I, M & 20.6 (7.0$\%$) & $\alpha$Her \nl
~~(V* V1302 Aql) \nl

IRAS 19343+2926 &  
19:36:17.5 & +29:32:50 &  5/16/91, I, B &  8.5CVF (10\%) &  $\alpha$Her \nl
~~(M 1$-$92) \nl

IRAS 19374+2359 & 
19:39:35.7 & +24:06:23 & 6/5/95, I, M & 9.8 (10.0$\%$) & $\alpha$Lyr \nl
 &  &  & 6/5/95, I, M & 12.5 (10.0$\%$) & $\alpha$Aql \nl
 &  &  & 6/5/95, I, M & 20.6 (7.0$\%$) &  $\alpha$Aql, $\alpha$Lyr \nl

IRAS 19386+0155 & 
19:41:08.2 & +02:02:31 & 11/10/95, I, M & 9.8 (10.0$\%$) & $\gamma$Aql \nl

IRAS 19454+2920 & 
19:47:24.3 & +29:28:16 & 6/6/95, I, M & 9.8 (10.0$\%$) & $\alpha$Lyr \nl

IRAS 19477+2401 & 
19:49:54.4 & +24:08:51 & 6/7/95, I, M & 9.8 (10.0$\%$) & $\alpha$Lyr \nl
 &  &  & 11/10/95, I, M & 12.5 (10.0$\%$) & $\gamma$Aql \nl

IRAS 19480+2504 & 
19:50:07.8 & +25:12:00 & 6/6/95, I, M & 9.8 (10.0$\%$) & $\alpha$Lyr \nl

IRAS 20000+3239 & 
20:01:59.4 & +32:47:32 & 6/3/95, I, M & 8.1CVF (1.8$\%$) & $\beta$Peg \nl
 &  &  & 6/3/95, I, M & 12.5 (10.0$\%$) & $\beta$Peg \nl
 &  &  & 6/6/95, I, M & 12.5 (10.0$\%$) & $\beta$Peg \nl

IRAS 20004+2955 & 
20:02:27.3 & +30:04:25 & 6/7/95, I, M & 9.8 (10.0$\%$) & $\alpha$Lyr \nl
~~(V* V1027 Cyg) \nl

IRAS 20028+3910 & 
20:04:36.1 & +39:18:44 & 11/10/95, I, M & 9.8 (10.0$\%$) &  $\epsilon$Cyg \nl
 &  &  & 11/10/95, I, M & 12.5 (10.0$\%$) & $\epsilon$Cyg \nl

IRAS 20043+2653 & 
20:06:22.9 & +27:02:11 & 11/22/95, U, M & 11.7 (10.0$\%$) & $\beta$Peg \nl
 
IRAS 20077$-$0625 & 
20:10:27.4 & $-$06:16:16 &  6/7/92, U, B &  12.2CVF (10\%) &  $\gamma$Aql  \nl
~~(IRC $-$10529) \nl

IRAS 21537+6435 & 
21:55:04 & +64:49:52 & 11/10/95, I, M & 12.5 (10.0$\%$) &  $\epsilon$Cyg, SAO19687 \nl

IRAS 22036+5306 & 
22:05:32.2 & +53:22:10 & 11/23/95, U, M & 8.8 (10.0$\%$) & $\beta$Peg \nl
 &  &  & 6/7/95, I, M & 12.5 (10.0$\%$) & $\beta$Peg \nl

IRAS 22142+5206 & 
22:16:10.3 & +52:21:32 & 11/10/95, I, M & 8.8 (10.0$\%$) & $\beta$Peg \nl
 &  &  & 11/10/95, I, M & 12.5 (10.0$\%$) & $\beta$Peg \nl

IRAS 22223+4327 & 
22:24:31 & +43:43:09 & 6/7/95, I, M & 12.5 (10.0$\%$) & $\beta$Peg \nl
 &  &  & 6/7/95, I, M & 18.0 (10.0$\%$) & $\beta$Peg \nl

IRAS 22480+6002 & 
22:49:59.1 & +60:17:55 & 11/10/95, I, M & 8.8 (10.0$\%$) &  $\beta$Peg, K star \nl
 &  &  & 11/10/95, I, M & 9.8 (10.0$\%$) &  $\beta$Peg, K star \nl
 &  &  & 11/10/95, I, M & 18.0 (10.0$\%$) & $\beta$Peg, K star \nl

IRAS 22574+6609 & 
22:59:18.4 & +66:25:48 & 11/10/95, I, M & 8.8 (10.0$\%$) &  $\beta$Peg, K star \nl
 &  &  & 6/7/95, I, M & 12.5 (10.0$\%$) & $\beta$Peg \nl

IRAS 23166+1655 & 
23:19:12.3  & +17:11:35 &  6/2/92, U, B &  8.5CVF (10\%) &  $\beta$Peg \nl
~~(AFGL 3068) &  &  &  12/13/92, U, B &  8.5CVF  (10\%) &  $\beta$Peg \nl
 &  &  &  12/13/92, U, B &  12.5CVF (10\%) &  $\beta$Peg \nl
 &  &  &  12/13/92, U, B &  11.3CVF (10\%) &  $\beta$Peg \nl
 
\tablebreak

IRAS 23304+6147 & 
23:32:44.5 & +62:03:58 & 6/5/95, I, M & 12.5 (10.0$\%$) & $\beta$Peg, $\alpha$Lyr \nl
 &  &  & 6/6/95, I, M & 12.5 (10.0$\%$) & $\beta$Peg \nl
 &  &  & 6/6/95, I, M & 20.6 (7.0$\%$) & $\beta$Peg \nl

IRAS 23321+6545 & 
23:34:21.8 & +66:01:53 & 11/7/95, I, M & 12.5 (10.0$\%$) &  $\beta$And, $\beta$Peg \nl
 &  &  & 11/10/95, I, M & 12.5 (10.0$\%$) &  $\beta$Peg, SAO19687 \nl

\tablenotetext{a}{Telescope $-$ I: IRTF, U: UKIRT}
\tablenotetext{b}{Camera $-$ M: MIRAC2, B: Berkcam}
\enddata
\end{deluxetable}

\begin{deluxetable}{lrrlcrlcrc}

\tablenum{3}
\tablewidth{0pt}
\tablecaption{Observed Fluxes and Sizes of Objects in MIRAC2 and Berkcam
Mid-IR Survey}

\tablehead{
\colhead{\iras} & \colhead{$\lambda$ [$\mu m$]} & \colhead{F [Jy]} & 
\multicolumn{3}{c}{\underline{~~~~~~~~~~~Source~~~~~~~~~~}} & 
\multicolumn{3}{c}{\underline{~~~~~~~~~~~~PSF~~~~~~~~~~~}} & 
\colhead{Standard} \\
\colhead{Identification}  &  &  & 
\colhead{Size\tablenotemark{a} [$^{\prime\prime}$]} & 
\colhead{EL\tablenotemark{b}} & 
\colhead{PA\tablenotemark{c}} & 
\colhead{Size\tablenotemark{a} [$^{\prime\prime}$]} & 
\colhead{EL\tablenotemark{b}} & 
\colhead{PA\tablenotemark{c}} & Star}
 
\startdata

00210+6221  & 12.5 & 33 & 1.10 & 0.10 & 72 & 1.10 & 0.06 & 97 &  $\beta$And \nl
  &    &    &    &    &    & 1.14 & 0.04 & 62 &  $\beta$And \nl
00470+6429  & 10.6 & 3 & 1.15 & 0.36 & 71 & 1.05 & 0.16 & 14 &  $\beta$And \nl
01144+6658  & 8.8 & 157 & 0.97 & 0.18 & 76 & 0.90 & 0.03 & 89 &  $\beta$And \nl
  &    &    &    &    &    & 0.86 & 0.04 & 43 &  $\beta$And \nl
  & 12.5 & 260 & 1.25 & 0.14 & 77 & 1.11 & 0.03 & 80 &  $\beta$And \nl
01174+6110  & 11.7 & 2 & 1.15 & 0.12 & 98 & 1.13 & 0.06 & 89 &  $\beta$And \nl
  &    &    &    &    &    & 1.12 & 0.10 & 85 &  $\beta$And \nl
01304+6211  & 9.8 & 73 & 1.06 & 0.08 & 88 & 1.10 & 0.13 & 82 &  $\beta$And \nl
  &    &    &    &    &    & 1.14 & 0.02 & 26 &  $\beta$And \nl
02143+5852  & 11.7 & \nodata & 0.96 & 0.06 & 165 & 0.98 & 0.11 & 80 &  $\beta$Peg \nl
02152+2822  & 11.3 & 162 & 1.00 & 0.15 & 174 & 1.20 & 0.09 & 163 &  $\beta$And \nl
02229+6208  & 8.8 & 30 & 1.10 & 0.07 & 72 & 0.90 & 0.06 & 49 &  $\beta$And \nl
  &    &    &    &    &    & 0.99 & 0.08 & 65 &  $\beta$And \nl
  & 9.8 & 41 & 1.11 & 0.05 & 51 & 0.97 & 0.02 & 26 &  $\beta$And \nl
  &    &    &    &    &    & 0.98 & 0.07 & 59 &  $\beta$And \nl
  & 11.7 & 87 & 1.27 & 0.10 & 66 & 1.05 & 0.07 & 58 &  $\beta$And \nl
  &    &    &    &    &    & 1.13 & 0.06 & 89 &  $\beta$And \nl
  & 12.5 & 90 & 1.29 & 0.07 & 63 & 1.14 & 0.04 & 62 &  $\beta$And \nl
  &    &    &    &    &    & 1.10 & 0.06 & 70 &  $\beta$And \nl
  & 20.6 & 86 & 1.93 & 0.21 & 95 & 1.67 & 0.06 & 56 &  $\alpha$Tau \nl
04386+5722  & 12.5 & 15 & 1.09 & 0.06 & 47 & 1.15 & 0.12 & 87 &  $\alpha$Tau \nl
  &    &    &    &    &    & 1.18 & 0.12 & 83 &  $\alpha$Tau \nl
05089+0459  & 11.7 & \nodata & 1.08 & 0.17 & 90 & 0.96 & 0.15 & 90 &  $\alpha$Tau \nl
05113+1347  & 11.7 & 4 & 1.12 & 0.05 & 38 & 1.07 & 0.06 & 51 &  $\alpha$Tau \nl
05341+0852  & 8.8 & 1 & 1.06 & 0.24 & 31 & 1.00 & 0.10 & 114 &  $\alpha$Tau \nl
  &    &    &    &    &    & 1.10 & 0.16 & 123 &  $\alpha$Tau \nl
  &  (11/7) 11.7  & 4 & 0.99 & 0.16 & 47 & 1.07 & 0.06 & 51 &  $\alpha$Tau \nl
  &  (11/23) 11.7  & \nodata & 0.92 & 0.09 & 96 & 0.89 & 0.20 & 97 &  $\alpha$Tau \nl
06176$-$1036  & 8.2 & 323 & 1.40 & 0.32 & 111 & 1.36 & 0.21 & 83 &  $\alpha$CMa \nl
  & 10.0 & 251 & 1.14 & 0.04 & 134 & 1.21 & 0.01 & 174 &  $\alpha$CMa \nl
  & 11.3 & 360 & 1.32 & 0.33 & 102 & 1.21 & 0.17 & 92 &  $\alpha$CMa \nl
06530$-$0213  & 12.5 & 12 & 1.18 & 0.06 & 47 & 1.15 & 0.12 & 87 &  $\alpha$Tau \nl
07239$-$3325  & 10.6 & \nodata & 0.74 & 0.09 & 98 & 0.98 & 0.26 & 89 &  $\alpha$CMi \nl
07284$-$0940  & 9.8 & \nodata & 1.05 & 0.30 & 88 & 1.01 & 0.31 & 88 &  $\alpha$CMi \nl
07331+0021  & 8.8 & \nodata & 0.97 & 0.25 & 89 & 0.89 & 0.20 & 97 &  $\alpha$Tau \nl
  &    &    &    &    &    & 1.01 & 0.24 & 85 &  $\alpha$CMa \nl
  & 20.6 & \nodata & 1.46 & 0.14 & 92 & 1.37 & 0.08 & 79 &  $\alpha$CMi \nl
07399$-$1435  & 8.8 & 25 & 1.66 & 0.25 & 43 & 1.15  & 0.14 & 21 & $\alpha$CMa  \nl
  & 9.8 & 20 & 1.96 & 0.20 & 22 & 0.98 & 0.11 & 116 & $\alpha$CMa  \nl
  & 11.7 & 30 & 1.61 & 0.25 & 26 & 1.05 & 0.04 & 20 & $\alpha$CMa  \nl
  & 18.0 & 204 & 2.10 & 0.25 & 13 & 1.30 & 0.04 & 20 & $\alpha$CMa  \nl
  & 20.6 & 290 & 2.28 & 0.19 & 9 & 1.75 & 0.27 & 0 & $\alpha$CMa  \nl
07430+1115  & 8.8 & \nodata & 0.93 & 0.18 & 96 & 0.89 & 0.20 & 97 &  $\alpha$Tau \nl
  & 11.7 & 6 & 1.10 & 0.21 & 39 & 1.07 & 0.23 & 50 &  $\alpha$Tau \nl
08005$-$2356  & 9.8 & \nodata   & 0.96 & 0.17 & 95 & 0.98 & 0.25 & 90 &  $\alpha$CMi \nl
  &    &    &    &    &    & 1.07 & 0.19 & 103 &  $\beta$Peg \nl
09032$-$3953  & 8.8 & \nodata & 1.00 & 0.18 & 107 & 0.89 & 0.13 & 90 &  $\alpha$CMi \nl
09371+1212  & 3.8 & \nodata & 1.07 & 0.08 & 144 & 1.29 & 0.06 & 93 &  $\alpha$CMa \nl
  & 10.6 & \nodata & 1.03 & 0.33 & 81 & 0.98 & 0.26 & 89 &  $\alpha$CMi \nl
10131+3049  & 11.5 & 2672 & 1.01 & 0.06 & 121 & 1.06 & 0.05 & 29 &  $\mu$UMa \nl
10158$-$2844  & 8.5 & 52 & 1.52 & 0.04 & 127 & 1.78 & 0.20 & 176 &  $\alpha$Boo \nl
  &    &    &    &    &    & 1.52 & 0.06 & 170 &  BS4104 \nl
  & 9.7 & 30 & 1.58 & 0.04 & 29 & 1.46 & 0.11 & 114 &  $\alpha$Boo \nl
  &    &    &    &    &    & 1.24 & 0.23 & 176 &  BS4104 \nl
11472$-$0800  & 9.8 & \nodata & 1.34 & 0.25 & 81 & 1.45 & 0.33 & 94 &  $\delta$Vir \nl
  &    &    &    &    &    & 1.03 & 0.15 & 103 &  $\mu$UMa \nl
15465+2818  & 12.5 & 24 & 1.21 & 0.05 & 81 & 1.15 & 0.32 & 99 &  $\alpha$Boo \nl
16342$-$3814  & 9.8 & 7 & 1.67 & 0.23 & 66 & 0.85 & 0.02 & 66 &  $\sigma$Lib \nl
  &    &    &    &    &    & 0.87 & 0.04 & 15 &  $\sigma$Lib \nl
  & 12.5 & 14 & 1.43 & 0.23 & 60 & 1.00 & 0.06 & 76 &  $\alpha$Boo \nl
  &    &    &    &    &    & 1.04 & 0.07 & 15 &  $\sigma$Lib \nl
  & 17.8 & 127 & 1.76 & 0.24 & 64 & 1.30 & 0.01 & 100 &  $\eta$Sgr \nl
  &    &    &    &    &    & 1.34 & 0.07 & 171 &  $\eta$Sgr \nl
  & 20.6 & 200 & 1.87 & 0.15 & 59 & 1.55 & 0.04 & 32 &  $\sigma$Lib \nl
  &    &    &    &    &    & 1.61 & 0.02 & 86 &  $\eta$Sgr \nl
16559$-$2957  & 12.5 & 11 & 1.02 & 0.04 & 34 & 1.01 & 0.01 & 69 &  $\eta$Sgr \nl
  &    &    &    &    &    & 1.03 & 0.05 & 42 &  $\eta$Sgr \nl
17028$-$1004  & 8.1 & 30 & 1.00 & 0.21 & 111 & 0.90 & 0.25 & 83 &  $\alpha$Her \nl
  & 8.5 & 41 & 0.89 & 0.22 & 105 & 0.92 & 0.12 & 75 &  $\alpha$Boo \nl
  &    &    &    &    &    & 0.88 & 0.30 & 88 &  $\alpha$Her \nl
  & 9.7 & 29 & 1.21 & 0.22 & 138 & 0.97 & 0.26 & 88 &  $\alpha$Her \nl
  & 12.6 & 44 & 1.31 & 0.17 & 166 & 1.20 & 0.06 & 126 &  $\alpha$Her \nl
  & 12.8 & 43 & 1.49 & 0.16 & 142 & 1.18 & 0.12 & 119 &  $\alpha$Her \nl
  &    &    &    &    &    & 1.22 & 0.21 & 127 &  $\alpha$Her \nl
17088$-$4221  & 12.5 & 62 & 1.10 & 0.11 & 13 & 1.04 & 0.04 & 104 &  $\eta$Sgr \nl
  &    &    &    &    &    & 1.01 & 0.01 & 69 &  $\eta$Sgr \nl
17106$-$3046  & 18.0 & 36 & 1.46 & 0.05 & 88 & 1.37 & 0.02 & 61 &  $\eta$Sgr \nl
  &    &    &    &    &    & 1.35 & 0.03 & 43 &  $\eta$Sgr \nl
17150$-$3224  & 8.1 & 15 & 1.22 & 0.13 & 68 & 0.87 & 0.04 & 96 &  $\eta$Sgr \nl
  &    &    &    &    &    & 1.06 & 0.13 & 70 &  $\alpha$Sco \nl
  & 9.8 & 24 & 1.26 & 0.11 & 121 & 0.98 & 0.05 & 172 &  $\alpha$Sco \nl
  &    &    &    &    &    & 0.97 & 0.07 & 172 &  $\alpha$Sco \nl
  & 12.5 & 74 & 1.27 & 0.11 & 72 & 1.03 & 0.05 & 42 &  $\eta$Sgr \nl
  &    &    &    &    &    & 1.16 & 0.10 & 62 &  $\eta$Sgr \nl
17245$-$3951  & 12.5 & 3 & 1.15 & 0.09 & 34 & 1.03 & 0.05 & 42 &  $\eta$Sgr \nl
  &    &    &    &    &    & 1.16 & 0.10 & 62 &  $\eta$Sgr \nl
17347$-$3139  & 9.8 & 8 & 1.33 & 0.20 & 138 & 0.96 & 0.04 & 165 &  $\alpha$Sco \nl
  &    &    &    &    &    & 0.96 & 0.04 & 158 &  $\alpha$Sco \nl
  & 12.5 & 28 & 1.20 & 0.13 & 145 & 1.05 & 0.03 & 125 &  $\alpha$Sco \nl
  &    &    &    &    &    & 1.00 & 0.02 & 173 &  $\alpha$Sco \nl
  & 20.6 & 102 & 1.62 & 0.04 & 136 & 1.61 & 0.01 & 90 &  $\alpha$Sco \nl
  &    &    &    &    &    & 1.55 & 0.01 & 94 &  $\eta$Sgr \nl
17436+5003  & 18.0 & 113 & 2.78 & 0.08 & 75 & 1.47 & 0.11 & 80 &  $\gamma$Dra \nl
  & 20.6 & 195 & 2.77 & 0.08 & 74 & 1.71 & 0.11 & 62 &  $\gamma$Dra \nl
\tablebreak
17441$-$2411  & 11.3 & 37 & 1.57 & 0.15 & 64 & 1.33 & 0.10 & 70 &  $\alpha$Her \nl
  &    &    &    &    &    & 1.29 & 0.10 & 82 &  $\eta$Sgr \nl
  & 12.5 & 51 & 1.46 & 0.13 & 58 & 1.16 & 0.09 & 52 &  $\eta$Sgr \nl
  &    &    &    &    &    & 1.27 & 0.06 & 86 &  $\eta$Sgr \nl
17534+2603  & 12.5 & 53 & 1.24 & 0.09 & 80 & 1.07 & 0.11 & 107 &  $\alpha$Her \nl
18071$-$1727  & 9.8 & 7 & 0.99 & 0.08 & 67 & 0.84 & 0.06 & 31 &  $\eta$Sgr \nl
  &    &    &    &    &    & 0.88 & 0.06 & 39 &  $\eta$Sgr \nl
  & 12.5 & 36 & 1.06 & 0.07 & 52 & 1.08 & 0.09 & 35 &  $\eta$Sgr \nl
18095+2704  & 8.5 & 12 & 1.00 & 0.01 & 45 & 1.03 & 0.11 & 159 &  $\alpha$Lyr \nl
  & 11.0 & 45 & 0.94 & 0.02 & 111 & 1.13 & 0.13 & 132 &  $\alpha$Lyr \nl
18184$-$1302  & 8.2 & 171 & 1.27 & 0.07 & 119 & 1.34 & 0.23 & 104 &  $\eta$Sgr \nl
  &    &    &    &    &    & 0.90 & 0.13 & 136 &  $\alpha$Lyr \nl
  & 12.2 & 305 & 1.27 & 0.05 & 30 & 1.02 & 0.23 & 170 &  $\alpha$Lyr \nl
18184$-$1623\tablenotemark{d}  & 8.8 & 41 & 18.4 & 0.39 & 22 & 1.83 & 0.29 & 148 &  $\sigma$Lib \nl
  &    &    &    &    &    & 1.82 & 0.24 & 39 &  $\sigma$Lib \nl
  &  (6/3)  12.5  & 64 & 18.2  & 0.26 & 36 & 1.00 & 0.05 & 85 &  $\beta$Peg \nl
  &    &    &    &    &    & 0.99 & 0.02 & 44 &  $\beta$Peg \nl
  &  (6/5)  12.5  & 59 & 17.9 & 0.32 & 22 & 1.95 & 0.53 & 150 &  $\eta$Sgr \nl
  & 20.6 & 302 & 20.5 & 0.43 & 23 & 1.80 & 0.03 & 111 &  $\alpha$Her \nl
18276$-$1431  & 9.8 & 13 & 1.13 & 0.12 & 34 & 0.94 & 0.15 & 48 &  $\alpha$Lyr \nl
  &    &    &    &    &    & 0.91 & 0.02 & 75 &  $\alpha$Sct\tablenotemark{e} \nl
  & 12.5 & 27 & 1.19 & 0.08 & 62 & 1.06 & 0.05 & 51 &  $\alpha$Lyr \nl
  &    &    &    &    &    & 1.13 & 0.13 & 82 &  $\alpha$Sct\tablenotemark{e} \nl
19075+0921  & 8.2 & 75 & 1.04 & 0.11 & 171 & 0.90 & 0.13 & 136 &  $\alpha$Lyr \nl
  & 12.2 & 149 & 1.03 & 0.08 & 6 & 1.02 & 0.23 & 170 &  $\alpha$Lyr \nl
19192+0922  & 12.5 & 229 & 1.13 & 0.08 & 131 & 1.05 & 0.06 & 108 &  $\gamma$Aql \nl
19244+1115  & 9.8 & 1987 & 1.31 & 0.16 & 45 & 1.18 & 0.09 & 58 &  $\alpha$Aql \nl
  & 20.6 & 3128 & 2.31 & 0.15 & 63 & 1.80 & 0.03 & 111 &  $\alpha$Her \nl
19343+2926  & 8.5 & 16 & 1.04 & 0.16 & 100 & 0.88 & 0.30 & 88 &  $\alpha$Her \nl
19374+2359  & 9.8 & 18 & 1.21 & 0.05 & 5 & 0.91 & 0.04 & 55 &  $\alpha$Lyr \nl
  & 12.5 & 28 & 1.27 & 0.07 & 17 & 1.05 & 0.11 & 63 &  $\alpha$Aql \nl
  & 20.6 & 176 & 1.83 & 0.05 & 57 & 1.57 & 0.02 & 116 &  $\alpha$Lyr \nl
  &    &    &    &    &    & 1.58 & 0.14 & 113 &  $\alpha$Aql \nl
19386+0155  & 9.8 & 10 & 1.19 & 0.20 & 86 & 1.20 & 0.24 & 87 &  $\gamma$Aql \nl
19454+2920  & 9.8 & 14 & 0.97 & 0.07 & 24 & 0.94 & 0.15 & 48 &  $\alpha$Lyr \nl
  &    &    &    &    &    & 0.92 & 0.09 & 89 &  $\alpha$Lyr \nl
19477+2401  & 9.8 & 6 & 0.97 & 0.08 & 74 & 1.00 & 0.13 & 71 &  $\alpha$Lyr \nl
  &    &    &    &    &    & 1.00 & 0.13 & 77 &  $\alpha$Lyr \nl
  & 12.5 & 13 & 1.35 & 0.17 & 88 & 1.33 & 0.17 & 90 &  $\gamma$Aql \nl
19480+2504  & 9.8 & 14 & 0.95 & 0.04 & 44 & 0.94 & 0.15 & 48 &  $\alpha$Lyr \nl
  &    &    &    &    &    & 0.92 & 0.09 & 89 &  $\alpha$Lyr \nl
20000+3239  & 8.1 & 5 & 1.13 & 0.21 & 85 & 1.01 & 0.08 & 70 &  $\beta$Peg \nl
  &  (6/3)  12.5  & 14 & 1.27 & 0.11 & 80 & 1.00 & 0.05 & 85 &  $\beta$Peg \nl
  &    &    &    &    &    & 0.99 & 0.02 & 44 &  $\beta$Peg \nl
  &  (6/6)  12.5  & 14 & 1.14 & 0.05 & 55 & 1.03 & 0.11 & 4 &  $\beta$Peg \nl
  &    &    &    &    &    & 1.00 & 0.06 & 11 &  $\beta$Peg \nl
20004+2955  & 9.8 & 54 & 0.93 & 0.06 & 80 & 0.96 & 0.10 & 68 &  $\alpha$Lyr \nl
  &    &    &    &    &    & 1.00 & 0.13 & 71 &  $\alpha$Lyr \nl
20028+3910  & 9.8 & 23 & 1.22 & 0.19 & 100 & 1.08 & 0.18 & 89 &  $\epsilon$Cyg \nl
  & 12.5 & 51 & 1.42 & 0.21 & 92 & 1.32 & 0.18 & 91 &  $\epsilon$Cyg \nl
  &    &    &    &    &    & 1.32 & 0.17 & 97 &  $\epsilon$Cyg \nl
20043+2653  & 11.7 & \nodata   & 1.05 & 0.08 & 30 & 1.00 & 0.12 & 103 &  $\beta$Peg \nl
  &    &    &    &    &    & 0.98 & 0.11 & 80 &  $\beta$Peg \nl
20077$-$0625  & 12.2 & 1089 & 1.13 & 0.03 & 172 & 1.27 & 0.09 & 162 &  $\gamma$Aql \nl
21537+6435  & 12.5 & 9 & 1.16 & 0.14 & 72 & 1.32 & 0.18 & 91 &  $\epsilon$Cyg \nl
  &    &    &    &    &    & 1.32 & 0.17 & 97 &  $\epsilon$Cyg \nl
  &    &    &    &    &    & 1.35 & 0.16 & 85 &  SAO19687\tablenotemark{e} \nl
22036+5306  & 12.5 & 6 & 1.20 & 0.14 & 72 & 1.33 & 0.33 & 88 &  $\beta$Peg \nl
  &    &    &    &    &    & 1.04 & 0.05 & 87 &  $\beta$Peg \nl
22142+5206  & 8.8 & 7 & 1.04 & 0.13 & 97 & 1.08 & 0.21 & 92 &  $\beta$Peg \nl
  &    &    &    &    &    & 0.90 & 0.10 & 75 &  $\beta$Peg \nl
  & 12.5 & 24 & 1.42 & 0.22 & 94 & 1.26 & 0.16 & 90 &  $\beta$Peg \nl
22223+4327  & 12.5 & 2 & 1.83 & 0.10 & 89 & 1.09 & 0.18 & 88 &  $\beta$Peg \nl
  &    &    &    &    &    & 1.01 & 0.06 & 86 &  $\beta$Peg \nl
  & 18.0 & 9 & 1.71 & 0.09 & 140 & 1.31 & 0.01 & 105 &  $\beta$Peg \nl
  &    &    &    &    &    & 1.33 & 0.03 & 78 &  $\beta$Peg \nl
22480+6002  & 8.8 & 91 & 0.99 & 0.15 & 80 & 0.90 & 0.10 & 75 &  $\beta$Peg \nl
  &    &    &    &    &    & 0.93 & 0.24 & 83 &  K star\tablenotemark{e} \nl
  & 9.8 & 195 & 1.10 & 0.16 & 73 & 0.89 & 0.04 & 19 &  $\beta$Peg \nl
  &    &    &    &    &    & 0.97 & 0.11 & 76 &  K star\tablenotemark{e} \nl
  & 18.0 & 184 & 1.76 & 0.15 & 88 & 1.49 & 0.05 & 110 &  $\beta$Peg \nl
  &    &    &    &    &    & 1.64 & 0.13 & 84 &  $\beta$Peg \nl
22574+6609  & 8.8 & 5 & 1.12 & 0.26 & 87 & 1.08 & 0.21 & 92 &  $\beta$Peg \nl
  &    &    &    &    &    & 0.90 & 0.10 & 75 &  $\beta$Peg \nl
  &    &    &    &    &    & 1.26 & 0.34 & 88 &  K star\tablenotemark{e} \nl
  & 12.5 & 9 & 1.21 & 0.15 & 89 & 1.04 & 0.05 & 87 &  $\beta$Peg \nl
  &    &    &    &    &    & 1.08 & 0.12 & 92 &  $\beta$Peg \nl
23166+1655  &  (6/2)  8.5  & 557 & 1.04 & 0.07 & 178 & 1.08 & 0.13 & 120 &  $\beta$Peg \nl
  &  (12/13) 8.5  & 284 & 1.09 & 0.12 & 129 & 0.81 & 0.08 & 132 &  $\beta$Peg \nl
  & 11.3 & 394 & 1.00 & 0.11 & 141 & 0.93 & 0.12 & 149 &  $\beta$Peg \nl
  & 12.5 & 435 & 1.12 & 0.07 & 108 & 1.18 & 0.06 & 125 &  $\beta$Peg \nl
23304+6147  &  (6/5) 12.5  & 13 & 1.45 & 0.23 & 83 & 1.10 & 0.08 & 80 &  $\alpha$Lyr \nl
  &    &    &    &    &    & 1.01 & 0.05 & 25 &  $\beta$Peg \nl
  &  (6/6) 12.5  & 12 & 1.37 & 0.16 & 82 & 1.00 & 0.06 & 11 &  $\beta$Peg \nl
  &    &    &    &    &    & 1.02 & 0.08 & 18 &  $\beta$Peg \nl
  & 20.6 & 58 & 1.62 & 0.03 & 42 & 1.56 & 0.04 & 31 &  $\beta$Peg \nl
23321+6545  &  (11/7) 12.5  & 15 & 1.18 & 0.06 & 90 & 1.10 & 0.06 & 97 &  $\beta$And \nl
  &    &    &    &    &    & 1.14 & 0.04 & 62 &  $\beta$And \nl
  &  (11/10) 12.5  & 18 & 1.15 & 0.10 & 87 & 1.26 & 0.16 & 90 &  $\beta$Peg \nl
  &    &    &    &    &    & 1.11 & 0.03 & 80 &  $\beta$And \nl
  &    &    &    &    &    & 1.35 & 0.16 & 85 &  SAO19687\tablenotemark{e}
\enddata
\tablenotetext{a}{The  major axis of the 50\% contour level.}
\tablenotetext{b}{$\mbox{EL}=1-b/a$, a and b being major and minor axis lengths.}
\tablenotetext{c}{Measured counter-clockwise from north in degrees.}
\tablenotetext{d}{F, Size, EL, and PA refer to those of the entire shell structure.}
\tablenotetext{e}{Not a flux calibrator.}
\end{deluxetable}

\def\ell{ .~.~.}
\begin{deluxetable}{lcccrrccc}

\tablenum{4a}
\tablewidth{0pt}
\tablecaption{Resolved Sources and Their Morphology}

\tablehead{
\colhead{\iras~ID} &
\colhead{$\lambda$ [$\mu$m]} & 
\colhead{Size\tablenotemark{a} [$^{\prime\prime}$]} & 
\colhead{PA}&
\colhead{Mid$-$IR} & 
\colhead{C/O\tablenotemark{b}} & 
\colhead{Type\tablenotemark{c}} & 
\colhead{Optical  } & 
\colhead{Ref.\tablenotemark{d}}\\
& & & &\colhead{ Morph. Class} &  & & \colhead{  Morph. Class}  }
 
\startdata

02229+6208  & 8.8 & 1.1 & \ell & unknown\tablenotemark{d} & C & PPN & star+bipolar & 7 \nl
  &  9.8 & 1.1 & \ell &  &  & \nl
  & 11.7 & 1.3 &\ell &  &  &   \nl
  &  12.5 & 1.3 &\ell &  &  &  \nl
  &  20.6 & 1.9 &\ell &  &  &  \nl
06176$-$1036  & 8.2 &$2.7\times 1.7$  &16  & core/elliptical & C/O & PPN & bipolar &  1\nl
~~(Red Rectangle)  &10.0 & $2\times 1.6$ &98  &  &  &  \nl
  & 11.3 & $2.6 \times 1.7$ & 18 &  &  &   \nl
07399$-$1435  & 8.8  & $2.8\times 2.2$  & 41& core/elliptical & O & Extreme AGB  & bipolar & 2 \nl
~~(OH 231.8+4.2)   &9.8 & $3.4\times 2.9$  & 22  &  &  & \nl
  & 11.7 & $2.9 \times 2.4$  & 29 &  &  &  \nl
  &  18.0 & $3.6\times 2.9$ & 17  &  &  & \nl
  &  20.6 & $3.8\times 3.2$ & 11  &  &  & \nl
16342$-$3814  &9.8 & $2.7\times 2.1$ & 68  & core/elliptical & O & Extreme AGB &  bipolar  & 7 \nl
  & 12.5 & $2.4 \times 1.8$ &  63  &  &  & \nl
  & 17.8 & $2.7 \times 2.1$ & 60  &  &  &  \nl
  & 20.6 & $2.9 \times 2.4$ &  67  &  &  & \nl
17028$-$1004  & 8.1 & $1.8\times 1.3$ & 97 & core/elliptical & O & Young PN & bipolar & 3 \nl
~~(M 2$-$9)  &  8.5 & $1.6\times 1.1$  & 101 &  &  & \nl
  & 9.7 & $2\times 1.3$ & 133  &  &  &  \nl
  &  12.6 & $2.5\times 2.2$ & 114  &  &  & \nl
  &  12.8 & $2.7 \times 2.2$ & 146  &  &  & \nl
17150$-$3224  & 8.1 & $2.1\times 1.7$  & 56 & core/elliptical & O & PPN & bipolar & 4,5,6,7\nl
  &  9.8 & $2.1 \times 1.8$  & 106 &  &  & \nl
  &  12.5 &$2.0 \times 1.8$  & 79 &  &  & \nl
17347$-$3139  & 9.8 & $2.1\times 1.8$ & 136 & core/elliptical & O & Young PN & unknown & \ell \nl
  &  12.5 & $1.8\times 1.6$ & 149  &  &  & \nl
  &  20.6 & $1.6 \times 1.2$ &  150 &  &  & \nl
17436+5003  &18.0 & $2.8\times 2.3$ & 99 & toroidal & O & PPN &  star+bipolar & 7 \nl
~~(HD 161796)  &  20.6 & $2.8\times 2.1$  &  95 & &  &  &\nl
17441$-$2411  & 11.3 & 1.6  & \ell & unknown\tablenotemark{d} & O & PPN & bipolar & 5,7 \nl
  &  12.5 & 1.5  &\ell &  &  & \nl
18184$-$1302  & 8.2 &1.3  & \ell & unknown\tablenotemark{d} & ? & Pre-MS & unknown & \ell \nl
~~(MWC 922)  & 12.2 & 1.3 & \ell  &  &  & \nl
18184$-$1623 & 8.8 & $18.4\times 11.2$ & 22 & toroidal & C & LBV/PPN & star+torus & 8  \nl
~~(HD 168625)  &   12.5 & $18.2\times 13.5$ & 36 & &  &  &   \nl
  & 20.6  & $20.5 \times 11.7$ &  23 &  &  &  \nl
18276$-$1431  & 9.8 & 1.1 &\ell & unknown\tablenotemark{d} & O & PPN & unknown & \ell  \nl
~~(OH 17.7$-$2.3)  & 12.5 & 1.2 &\ell   &  &  & \nl
19244+1115  & 9.8 &  $1.3\times 1.2$ & 58 & core/elliptical & O & SG & bipolar & 9 \nl
~~(IRC $+$10420)  & 20.6 & $2.3 \times 1.9$ &  75  &  &  & \nl
19374+2359  & 9.8 & $1.2\times 1.1$  & 34 & core/elliptical & O & PPN & star+round & 10 \nl
  &12.5 & $1.3\times 1.2$ &  50 &   &  &  &  \nl
  & 20.6 & $1.8 \times 1.7$ &  51  &  &  & \nl
20000+3239  & 8.1 & 1.1 & \ell & unknown\tablenotemark{d} & C & PPN & star+round & 10  \nl
  &   12.5 & 1.1 &\ell &  &  &   \nl
\tablebreak
22223+4327  &12.5 & $1.9\times 1.7$ & 108 & toroidal & C & PPN & unknown & \ell \nl
  & 18.0 & $1.9 \times 1.8$  & 152  &  &  & \nl
23304+6147  &   12.5 & 1.5 & \ell & unknown\tablenotemark{d} &  C & PPN  & star+round & 10 \nl
  &20.6 & 1.6 & \ell  &  &  &  \nl

\tablenotetext{a}{For extended sources, the size is  major$\times$minor axes of
the 20\% contour level.
For marginally resolved sources (i.e. morph. class unknown), the size is
the major axis of the 50\% contour level which is, of course, an upper
limit to the actual size of the source.}
\tablenotetext{b}{
C = carbon rich; O = oxygen rich; C/O = both carbon and oxygen rich
}
\tablenotetext{c}{
PPN: Proto-Planetary Nebulae, 
PN: Planetary Nebulae, 
AGB:  Asymptotic Giant Branch star, 
LBV: luminous blue variable,
SG: Super Giant, and 
MS: Main Sequence}
\tablenotetext{d}{1. Cohen et al. (1975); 2. Kastner et al. (1992);
3. Trammell, Goodrich, \& Dinerstein (1995); 4. Hu et al. (1993b);
5. Kwok et al. (1996); 6. Kwok, Su \& Hrivnak (1998); 7. Ueta, Meixner \&
Bobrowsky (1998);
8. Hutsemekers et al. (1994); 9. Humphreys et al. (1993); 10.  Hrivnak et al. 
(1999) }
\tablenotetext{d}{``Unknown'' is assigned to sources which are only marginally
resolved.}

\enddata
\end{deluxetable}

\def\ell{ .~.~.}
\begin{deluxetable}{llcccccc}

\tablenum{4b}
\tablewidth{0pt}
\tablecaption{Mid--Infrared  Images from the Literature}

\tablehead{
\colhead{\iras /IR ID} &
\colhead{alternate name} & 
\colhead{Mid-IR} & 
\colhead{Ref.\tablenotemark{a}}&
\colhead{C/O\tablenotemark{b}} & 
\colhead{Type\tablenotemark{c}} & 
\colhead{Optical} & 
\colhead{Ref.\tablenotemark{d}} \\
& & \colhead{Morph. Class} & & & & \colhead{Morph. Class}
}

\startdata

04296+3429 & \ell & unresolved & 2 & C & PPN & star+multipolar &  c \nl
04395+3601 & AFGL 618  & core/elliptical & 6 & C & Young PN & bipolar & a \nl
06176$-$1036 & AFGL 915 & core/elliptical & 6 & C/O & PPN &bipolar & b  \nl
07134+1005 & HD56126  & toroidal& 2,3 & C & PPN & star+bipolar &c \nl
17436+5003 & HD 161796 & toroidal & 7&  O & PPN & star+bipolar & c \nl 
19114+0002 & HD179821,AFGL 2343  & toroidal & 1 & O & PPN/SG  & star+multipolar & c \nl
19500-1709 & HD187885 & unknown& 2 & C & PPN & unknown & \ell \nl
AFGL 2688   &  Egg Nebula & core/elliptical & 5,6 & C & PPN  & bipolar & d \nl
21282+5050 & \ell & toroidal & 2,4 & C & Young PN & star+bipolar & e \nl
22272+5435 & HD235858 & toroidal& 2,3  & C & PPN & star+multipolar & c \nl

\tablenotetext{a}{  
1. Hawkins et al. (1995);
2. Meixner et al. (1997);
3. Dayal et al. (1998);
4. Meixner et al. (1993b);
5. Skinner et al. (1997);
6. Hora et al. (1996);
7. Skinner et al. (1994)
}

\tablenotetext{b}{
C = carbon rich; O = oxygen rich; C/O = both carbon and oxygen rich
}

\tablenotetext{c}{
PPN: Proto-Planetary Nebulae; 
PN: Planetary Nebulae
}
\tablenotetext{d}{
a. Westbrook et al. (1975);
b. Cohen et al. (1975);
c. Ueta, Meixner, Bobrowsky (1998);
d.  Ney et al. (1975); 
e. Meixner (1993)
}

\enddata
\end{deluxetable}